\documentclass{article}

\usepackage{arxiv}

\usepackage[utf8]{inputenc} 
\usepackage[T1]{fontenc}    
\usepackage{hyperref}       
\usepackage{url}            
\usepackage{booktabs}       
\usepackage{amsfonts}       
\usepackage{nicefrac}       
\usepackage{microtype}      
\usepackage{cleveref}       
\usepackage{lipsum}         
\usepackage{graphicx}
\usepackage{natbib}
\usepackage{doi}
\usepackage{caption}
\usepackage{subcaption}
\usepackage{xcolor}
\newcommand{\blue}[1]{\textcolor{black}{#1}}
\usepackage{multirow}
\usepackage{subcaption}

\title{PBSCR: The Piano Bootleg Score Composer Recognition Dataset}

\newif\ifuniqueAffiliation
\uniqueAffiliationtrue

\ifuniqueAffiliation 
\author{Arhan Jain\thanks{Equal contribution} \\
	University of Washington\\
	Seattle, WA 98195 \\
	\texttt{arhanj@cs.washington.edu} \\
	\And
	Alec Bunn$^*$ \\
	University of Washington\\
	Seattle, WA 98195 \\
	\texttt{abunn2@uw.edu} \\
	\And
	Austin Pham \\
	Columbia University \\
	New York, NY 10027\\
	\texttt{ap4460@columbia.edu} \\
	\And
	TJ Tsai \\
	Harvey Mudd College \\
	Claremont, CA 91711\\
	\texttt{ttsai@g.hmc.edu} \\
}
\else
\usepackage{authblk}

\newbox{\orcid}\sbox{\orcid}{\includegraphics[scale=0.06]{orcid.pdf}} 
\author[1]{%
	\href{https://orcid.org/0000-0000-0000-0000}{\usebox{\orcid}\hspace{1mm}David S.~Hippocampus\thanks{\texttt{hippo@cs.cranberry-lemon.edu}}}%
}
\author[1,2]{%
	\href{https://orcid.org/0000-0000-0000-0000}{\usebox{\orcid}\hspace{1mm}Elias D.~Striatum\thanks{\texttt{stariate@ee.mount-sheikh.edu}}}%
}
\affil[1]{University of Washington, Seattle, WA 98195}
\affil[2]{Harvey Mudd College, Claremont, CA 91711}
\fi

\renewcommand{\shorttitle}{\textit{arXiv} Template}

\hypersetup{
	pdftitle={PBSCR: The Piano Bootleg Score Composer Recognition Dataset},
	pdfsubject={cs.SD, cs.MM},
	pdfauthor={Arhan Jain, Alec Bunn, Austin Pham, TJ Tsai},
	pdfkeywords={composer recognition, sheet music, dataset, piano},
}

\begin{document}
	\maketitle
	
	\begin{abstract}
This article motivates, describes, and presents the PBSCR dataset for studying composer recognition of classical piano music.  \blue{Our goal was to design a dataset that facilitates large-scale research on composer recognition that is suitable for modern architectures and training practices.  To achieve this goal, we utilize the abundance of sheet music images and rich metadata on IMSLP, use a previously proposed feature representation called a bootleg score to encode the location of noteheads relative to staff lines, and present the data in an extremely simple format (2D binary images) to encourage rapid exploration and iteration.}  The dataset itself contains 40,000 62x64 bootleg score images for a 9-class recognition task, 100,000 62x64 bootleg score images for a 100-class recognition task, and 29,310 unlabeled variable-length bootleg score images for pretraining.  The labeled data is presented in a form that mirrors MNIST images, in order to make it extremely easy to visualize, manipulate, and train models in an efficient manner.  \blue{We include relevant information to connect each bootleg score image with its underlying raw sheet music image, and we scrape, organize, and compile metadata from IMSLP on all piano works to facilitate multimodal research and allow for convenient linking to other datasets.  We release baseline results in a supervised and low-shot setting for future works to compare against, and} we discuss open research questions that the PBSCR data is especially well suited to facilitate research on.

	\end{abstract}
	
	
\section{Introduction}
\label{sec:intro}

Composer recognition is the task of predicting the composer of an unseen piece or fragment of music.  Similar to other music classification tasks like genre recognition, emotion/mood detection, and music tagging, a composer recognition system allows one to describe stylistic elements in a piece of music and is thus useful in music recommendation and organization of digital music collections.  Unlike these other tasks, however, composer recognition characterizes the style in a self-referential manner, using similarities to known composers rather than subjective, human-generated labels.  In an age where data is often the limiting factor in training powerful and expressive models, composer recognition offers several desirable characteristics as a music classification task: it has objective (rather than subjective) ground truth labels, it does not require expensive human labeling, and the task difficulty can be adjusted by including more or fewer composers.  The task also offers a testbed for developing deeper musicological insights into the stylistic differences between composers, eras, and historical movements.  Our goal in this work is to design a dataset that facilitates large-scale research on composer recognition that is suitable for modern architectures and practices, and that presents a large database of sheet music images and metadata in a form that is accessible and easy to use.

While composer recognition has received a lot of interest in the last 20 years, previous work has been hindered by major data constraints.  Most previous works have focused on symbolic representations of music like MIDI or **kern where note and duration information are explicitly encoded.  One significant drawback of using these representations is that they are less common than audio or sheet music, and are thus limited to smaller scale datasets of varying quality collected from various websites.  The Lakh MIDI dataset \citep{raffel2016learning} has assembled a large amount of MIDI data from various websites but does not come with metadata.  The lack of large-scale datasets has become a major bottleneck in recent years as models have become dependent on large amounts of data for pretraining and finetuning.  Audio data is available in abundance but has been limited by copyright restrictions, which has hindered the systematic collection and organization of large, open datasets for composer recognition.  Sheet music is available in abundance with high-quality metadata and permissive licenses on IMSLP\footnote{https://imslp.org}, but it struggles with a different problem: an inconvenient format.  Raw sheet music images do not directly encode note information, and current optical music recognition (OMR) capabilities are unreliable on scanned sheet music of varying quality.

The GiantMIDI-Piano dataset \citep{kong2022giantmidi} proposes one way to address these constraints: it downloads Youtube recordings of solo piano works, processes the recordings with an automatic music transcription (AMT) system, and releases only the estimated MIDI transcriptions.  This has several benefits: it utilizes the fact that audio recordings are plentiful, avoids copyright issues by not releasing the audio, and provides the music in symbolic form.  But it also has several drawbacks when used to study composer recognition: the note transcriptions are noisy estimates, the metadata is not entirely reliable (since it relies on a Youtube search based on composer name and piece title), and the audio conflates performance and compositional aspects.

This article proposes an alternative way to address these data constraints: it uses solo piano sheet music images from IMSLP, extracts a previously proposed feature representation called a bootleg score, and presents the features in a compact and convenient format (binary 2D images).  The bootleg score \citep{yang2019midi} uses classical computer vision techniques to detect noteheads and encodes their locations relative to the staff lines.  It can be thought of as a redacted onset-only piano roll where duration and accidental information have been discarded.  This approach has several benefits: it utilizes the fact that classical sheet music is plentiful, it has rich and reliable metadata from IMSLP, and it provides the data in a symbolic format.  Its drawbacks are that the notehead detection and localization are noisy estimates, the bootleg score only encodes a selected set of musically relevant information, and the format is less commonly used.  This approach can be thought of as a companion and complement to the GiantMIDI-Piano dataset, where the focus is on creating a useful research dataset from sheet music images rather than Youtube recordings.

The proposed dataset can be useful in a number of ways.  First and foremost, it is designed to facilitate progress on composer recognition by providing the largest-scale benchmark to date.  Even though it only contains selected information about notes, it allows us to study the problem at a larger scale with modern architectures and gain a deeper understanding of musical aspects that are distinctive about individual composers (e.g.,~texture, melodic contours, etc).  Second, the simplicity of the dataset (2D images patterned after MNIST) can enable rapid experimentation on questions of broader interest to the MIR community, such as designing effective music representations or data augmentation strategies.  Third, the rich (and reliable) metadata associated with the bootleg scores can serve as a foundation for multimodal research, linking together various data sources like sheet music, audio recordings, MIDI files, composer and piece metadata, relevant wikipedia pages, and All Music Guide descriptions (all of which are linked on IMSLP).  For example, \cite{yang2021composer} demonstrate the feasibility of cross-modal transfer learning, in which a model is trained only on (sheet music) bootleg scores and used to perform composer classification of audio recordings.  Fourth, the dataset can be used to link and de-anonymize data from other datasets.  For example, \cite{yang2021piano} used bootleg scores to identify matches between IMSLP sheet music and the Lakh MIDI dataset \citep{raffel2016learning}.  Similar techniques could be used to verify and clean datasets like GiantMIDI-Piano \citep{kong2022giantmidi} that have unreliable metadata.  Fifth, the dataset can be used to study large-scale retrieval problems like piece or passage identification of audio, MIDI, or sheet music images \citep{yang2022large}.

The main contribution of this article is to describe, introduce, and motivate the Piano Bootleg Score Composer Recognition (PBSCR) dataset.\footnote{This article is a journal extension to \cite{tsai2020composer} with an exclusive focus on the dataset (and not the techniques).  This article focuses on expanding, improving, and making this previous dataset as easy to use as possible.  The novel contributions in this journal include: (a) going to significant lengths to clean up the large, unlabeled dataset for pretraining by removing non-music filler pages (Section 3), (b) expanding the labeled dataset from 9 composers to 100 (Section 4), (c) adding metadata information from IMSLP to facilitate multimodal research and allow for convenient linking to other datasets (Section 3.3), (d) presenting a new set of composer classification results using the updated dataset (Section 5), and (e) offering a comprehensive discussion of the context (Section 2), data leakage issues (Section 4.3), and research questions relevant to this dataset (Section 6).}  This dataset was designed to facilitate research on composer recognition with a focus on size, diversity, and ease of use.  Our guiding motto was to design a composer recognition dataset that is ``as accessible as MNIST and as challenging as ImageNet.''  Thus, our goal was to design a dataset that presents a challenging task but is as compact, lightweight, and easy to work with and visualize as MNIST images.  To maximize data quantity while ensuring data simplicity and ease of use, we consider piano sheet music images from IMSLP and use a previously proposed sheet music feature representation called a bootleg score \citep{yang2019midi} to encode the locations of noteheads relative to staff lines.  The dataset consists of three parts: a labeled set of 40,000 62x64 piano bootleg score images for a 9-class composer recognition task, a labeled set of 100,000 62x64 piano bootleg score images for a 100-class composer recognition task, and a large unlabeled set of variable-length piano bootleg scores in IMSLP for self-supervised learning.  For each labeled piano bootleg score, the dataset includes information to allow researchers to access the raw sheet music images from which the bootleg score fragment was taken.  For both labeled and unlabeled data, we also scrape, organize, and compile metadata from IMSLP about each work.  We release a set of baseline systems and results for researchers to compare against in future works.  In addition, we discuss several research tasks and open research questions that the PBSCR dataset is especially well suited to study.  This discussion lays out interesting directions and potential roadmaps for future work.  The dataset and code for this project can be found at \url{https://github.com/HMC-MIR/PBSCR}.

\section{Background}
\label{sec:background}

In this section we provide background about previous work and datasets used to study the composer recognition task.

\subsection{Previous Work}
\label{subsec:prevWork}

In this subsection we provide a brief overview of previous methods in composer recognition, and we describe how recent methods motivate a need for larger datasets.  Previous work can be divided into two time periods: classical machine learning and deep learning.

Classical machine learning methods were the dominant approach between roughly 2000 and 2015.  Most methods for composer classification from this era fall into one of two categories.  The first category is to define a set of manually designed features, and then feed the features into a standard classifier.  Some examples of features include absolute and relative values of pitch and duration \citep{pape2008democratic}, global statistics on pitch intervals and durations \citep{goienetxea2018use}, chroma-based features \citep{anan2012polyphonic}, n-gram statistics on intervals and durations \citep{hajj2018automated}, high-level musicological features like detecting 9-8 suspensions \citep{brinkman2016musical} or detecting sonata form \citep{kempfert2020does}, and using standardized feature sets \citep{herremans2016composer} like the jSymbolic toolbox \citep{mckay2006jsymbolic}.  The second category of classical machine learning approaches is to train a sequence-based model.  The two most common sequence-based models from this era are $N$-gram models (e.g., \citep{hontanilla2013modeling}) and Markov chains (e.g., \citep{hedges2014predicting}).  In this approach, a sequence-based model is trained on each composer of interest, and test sequences are classified by selecting the model that has the highest likelihood.

Deep learning-based approaches have increasingly become the dominant paradigm since around 2015.  Most methods in this era fall into one of two categories.

The first category represents the underlying music information as a continuous signal.  Common representations include mel spectrogram and MFCC features for audio (e.g., \cite{micchi2018neural, kher2022thesis}), a piano roll-like matrix or tensor specifying note events for symbolic music (e.g., \cite{verma2019midi, velarde2018convolution}), and 2D images for sheet music \citep{walwadkar2022compldnet}.  Given an input represented as a continuous signal, various neural network architectures have been explored including Convolutional Neural Network (CNN) architectures (e.g., \cite{walwadkar2022compldnet, deepaisarn2022visual, kim2020deep}), Convolutional Recurrent Neural Networks (CRNNs) \citep{kong2020large}, and Long Short-Term Memory (LSTM) models \citep{micchi2018neural, kher2022thesis}.

The second category represents the underlying music information as a sequence of discrete tokens.  Some methods for forming discrete tokens from music data include: converting piano roll-like data into binary text \citep{takamoto2018feature} or sequence of characters \citep{yang2021composer}, considering (note,duration) tuples \citep{deepaisarn2023nlp}, or using a REMI \citep{huang2020pop} or compound word representation \citep{hsiao2021compound}.  Once the data is represented as a sequence of discrete tokens, a variety of Transformer architectures have been used to model the data \citep{li2023fine, chou2021midibert, yang2021deeper, tsai2020composer}.  One benefit of this approach is the ability to pretrain models on unlabeled data in a self-supervised manner.  These methods are data-hungry and provide a strong incentive to construct benchmarks that contain large amounts of data.

\subsection{Previous Datasets}
\label{subsec:datasets}

In this section we describe the landscape of datasets used to study the composer classification task.  This provides historical context for understanding the contribution of the PBSCR dataset.

Table \ref{tab:dataComparison} provides an overview of recent works on composer classification and the datasets used in these studies.  From left to right, the columns indicate the paper (author and year published), number of composers in the classification task, original source data type (symbolic, audio, sheet music), preprocessed data format (i.e., after any data preprocessing), and dataset size.  For dataset size, numbers in parentheses indicate unlabeled data for pretraining.  Entries in the table have been sorted by publication year, and the last entry in the table corresponds to the proposed PBSCR dataset.

\begin{table*}
	\centering
	\begin{tabular}{lcccc}
		\toprule
		Paper & Composers & \blue{Original Source} & \blue{Preprocessed} & Data Size \\
		& & \blue{Data Type} & \blue{Data Format} &  \\ \midrule
		\cite{wolkowicz2013evaluation} & 5 & symbolic & \blue{MIDI} & 251 \blue{pieces} \\
		\cite{hontanilla2013modeling} & 5 & symbolic & \blue{MIDI} & 274 \blue{movements} \\
		\cite{herlands2014machine} & 2 & symbolic & \blue{MIDI} & 74 \blue{movements} \\
		\cite{hedges2014predicting} & 9 & symbolic & \blue{chords} & 5700 lead sheets\\
		\cite{herremans2015classification} & 3 & symbolic & \blue{MIDI} & 1045 \blue{pieces} \\
		\cite{saboo2015composer} & 2 & symbolic & \blue{museData, kern} & 366 \blue{pieces}  \\
		\cite{brinkman2016musical} & 6 & symbolic & \blue{no info} & no info \\
		\cite{velarde2016composer} & 2 & symbolic & \blue{kern} & 107 \blue{movements}  \\
		\cite{herremans2016composer} & 3 & symbolic & \blue{MIDI} & 1045 \blue{movements} \\
		\cite{shuvaev2017representations} & 31 & audio & \blue{audio} & 62 hrs \\
		\cite{sadeghian2017classification} & 3 & symbolic & \blue{MIDI} & 417 \blue{sonatas}  \\
		\cite{takamoto2018feature} & 5 & symbolic & \blue{MIDI} & 75 \blue{pieces}  \\ 
		\cite{hajj2018automated} & 9 & symbolic & \blue{MIDI} & 1197 \blue{pieces}  \\
		\cite{velarde2018convolution} & 5 & symbolic & \blue{MIDI}, audio (synthesized) & 207 \blue{movements} \\
		\cite{micchi2018neural} & 6 & audio & \blue{audio} & 320 recordings \\
		\cite{goienetxea2018use} & 5 & symbolic & \blue{kern} & 1586 \blue{pieces}  \\
		\cite{verma2019midi} & 19 & symbolic & \blue{kern} & 2500 \blue{pieces}  \\
		\cite{costa2019dodecaphonic} & 3 & symbolic & \blue{no info} & 10 \blue{pieces}  \\
		\cite{kim2020deep} & 13 & symbolic & \blue{MIDI} & 505 \blue{pieces}  \\
		\cite{kong2020large} & 100 & audio & \blue{MIDI (transcribed)} & 10854 \blue{pieces} \\
		\cite{revathi2020robust} & 4 & audio & \blue{audio} & 40 \blue{pieces}  \\
		\cite{kempfert2020does} & 2 & symbolic & \blue{kern} & 285 \blue{movements}  \\
		\cite{chou2021midibert} & 8 & symbolic & \blue{MIDI} & 411 \blue{pieces}  \\
		\cite{yang2021composer} & 9 & sheet music & \blue{bootleg score} & 787 \blue{works} (29310 \blue{works}) \\
		\cite{walwadkar2022compldnet} & 9 & sheet music & \blue{image, bootleg score} & 32k images \\
		\cite{deepaisarn2022visual} & 5 & symbolic & \blue{MIDI} & 809 \blue{pieces}  \\
		\cite{kher2022thesis} & 11 & symbolic & \blue{MIDI, audio (synthesized)} & 110 \blue{pieces}  \\
		\cite{foscarin2022concept} & \blue{13} & \blue{symbolic} & \blue{MIDI} & \blue{667 pieces} \\
		\cite{li2023fine} & 8 & audio & \blue{MIDI (transcribed)} & 411 \blue{pieces}  \\
		\cite{deepaisarn2023nlp} & 5 & symbolic & \blue{MIDI} & 809 \blue{pieces}  \\
		\cite{simonetta2023optimizing} & 7 & symbolic & \blue{MIDI} & 211 \blue{pieces}  \\
		\cite{zhang2023symbolic} & \blue{9} & \blue{symbolic} & \blue{MIDI, MusicXML} & \blue{415 scores} \\ \midrule
		PBSCR & 100 & sheet music & \blue{bootleg score} & 4997 \blue{works} (29310 \blue{works}) \\
		\bottomrule
	\end{tabular}
	\caption{\blue{Overview of recent works on composer classification and the datasets used in these studies.  The third column indicates whether the original source data is symbolic, audio, or sheet music images.  The fourth column indicates the format of the data after any data format conversions or preprocessing.  The fifth column indicates the size of the dataset, where numbers in parentheses indicate unlabeled files for pretraining.  For papers that use multiple datasets, we have only indicated the largest.}}
	\label{tab:dataComparison}
\end{table*}

\blue{There are three things to notice about the landscape of previous datasets described in Table \ref{tab:dataComparison}.  First, most previous works consider a small number of composers.  For example, only 6 out of the 32 previous works shown in Table \ref{tab:dataComparison} consider more than 10 composers.  Second, most previous works consider a relatively small amount of data by modern standards.  It is difficult to compare dataset sizes directly since previous works report sizes in different ways, including number of movements/pieces, total audio duration, and number of sheet music images.  Nonetheless, comparing entries by the number of pieces in the labeled dataset (the most common metric), we can see that most works consider on the order of hundreds of pieces.  Third, the vast majority of previous work focuses on symbolic music formats.  As a practical matter, the choice to use symbolic music as source data limits the size and diversity of datasets, since symbolic music data is less plentiful than sheet music or audio.}

\blue{It is useful to point out how the PBSCR dataset fits into this data landscape.  It is distinctive in three ways.  First, it has the highest number of composers (100, tied with \cite{kong2020large}).  As mentioned above, this is much higher than most previous work, so it poses a more challenging classification task.  Second, the PBSCR dataset is among the largest in size.  In particular, it is one of the only datasets in Table \ref{tab:dataComparison} that comes with a large unlabeled dataset for pretraining.  Given the shift in recent years towards pretraining models on unlabeled data in a self-supervised manner, this provides an essential resource for supporting the development of competitive models.  Based on the number of works in both the labeled (4997) and unlabeled (29310) datasets, the PBSCR dataset is almost certainly the largest in terms of total dataset size.  Third, it is one of only a few works \citep{walwadkar2022compldnet, tsai2020composer, yang2021composer} that uses sheet music images as source data.  By using a bootleg score representation, the PBSCR dataset maintains the advantage of plentiful sheet music data (on IMSLP) while presenting the data in an extremely compact and simple form (binary 2D images).}

\blue{Given this context,} we can reasonably make the following claim: \blue{the PBSCR dataset presents the most challenging classification task} (based on the number of composer classes), has the largest and most diverse set of data available (based on number of pieces and composers), and has the simplest and most accessible data format (2D binary images).

It is useful to note that the PBSCR dataset has a very different philosophy from most previous works in composer recognition.  Whereas previous approaches require full symbolic music information and accept the consequence of limited data size \& diversity, the PBSCR dataset first requires that the dataset be large, open, diverse, and easy to work with and accepts the consequence of a noisy, selective feature representation.  By using the bootleg score representation, we construct a dataset that is as easy to work with as MNIST data and can facilitate rapid exploration and iteration.

\section{Dataset Preparation: Unlabeled Data}
\label{sec:dataUnlabeled}

\blue{The PBSCR Dataset consists of three parts: a large set of unlabeled piano bootleg scores for pretraining (Section~\ref{subsec:imslpData}), a set of labeled data for a 9-class composer recognition task (Section~\ref{subsec:data9way}), and a set of labeled data for a 100-class composer recognition task (Section~\ref{subsec:data100way}).}  In this section, we describe the preparation of the unlabeled dataset, which we refer to as the IMSLP Piano Bootleg Scores Data (v1.1).  The labeled datasets will be described in Section~\ref{sec:dataLabeled}.  \blue{The novel contributions of the v1.1 dataset are (a) identifying and removing non-music filler pages from the v1.0 dataset (Section~\ref{subsec:identifyFiller}) and (b) scraping, organizing, and including metadata from IMSLP on all works to facilitate multimodal research and allow for convenient linking to other datasets (Section~\ref{subsec:imslpMetadata}).}

\subsection{IMSLP Piano Bootleg Scores v1.1}
\label{subsec:imslpData}

\blue{The IMSLP piano bootleg scores repository (v1.0) was introduced in \cite{yang2020camera} for a sheet music identification task, and first used for composer classification in \cite{tsai2020composer}}.  At a high level, this repository contains the bootleg scores for all solo piano works in IMSLP.  Below, we describe the history of its construction, as well as the new steps that were taken to clean up the data (v1.1).

The first step in creating this repository was to download IMSLP sheet music.  \blue{The IMSLP website provides a list of composers, a list of works for each composer, and a webpage for each work that contains audio recordings, sheet music, and metadata.  For each composer, we iterated through all of their works and downloaded all PDF sheet music scores and associated metadata.  The scraping and downloading took over a month to complete, resulting in a set of $420,271$ PDF files from $164,248$ composers that was 1.2 terabytes in size.}

The second step was to filter the full dataset by instrumentation tag label in order to identify a list of solo piano \blue{works}.  After filtering, the dataset contained $29,310$ \blue{works}, $31,384$ PDFs, and $374,758$ individual pages.  \blue{Note that a work may contain several PDF versions on the IMSLP website.}  All of the remaining steps below were applied only to this filtered dataset.

The third step is to convert each PDF into a sequence of PNG images.  We perform the decoding at 300 dpi, and then resize the image to have a fixed width of 2550 pixels while preserving the aspect ratio.  This resizing step is necessary to appropriately handle the extremely large range of image sizes in IMSLP.

\begin{figure}
	\centering
	\includegraphics[width=0.5\linewidth]{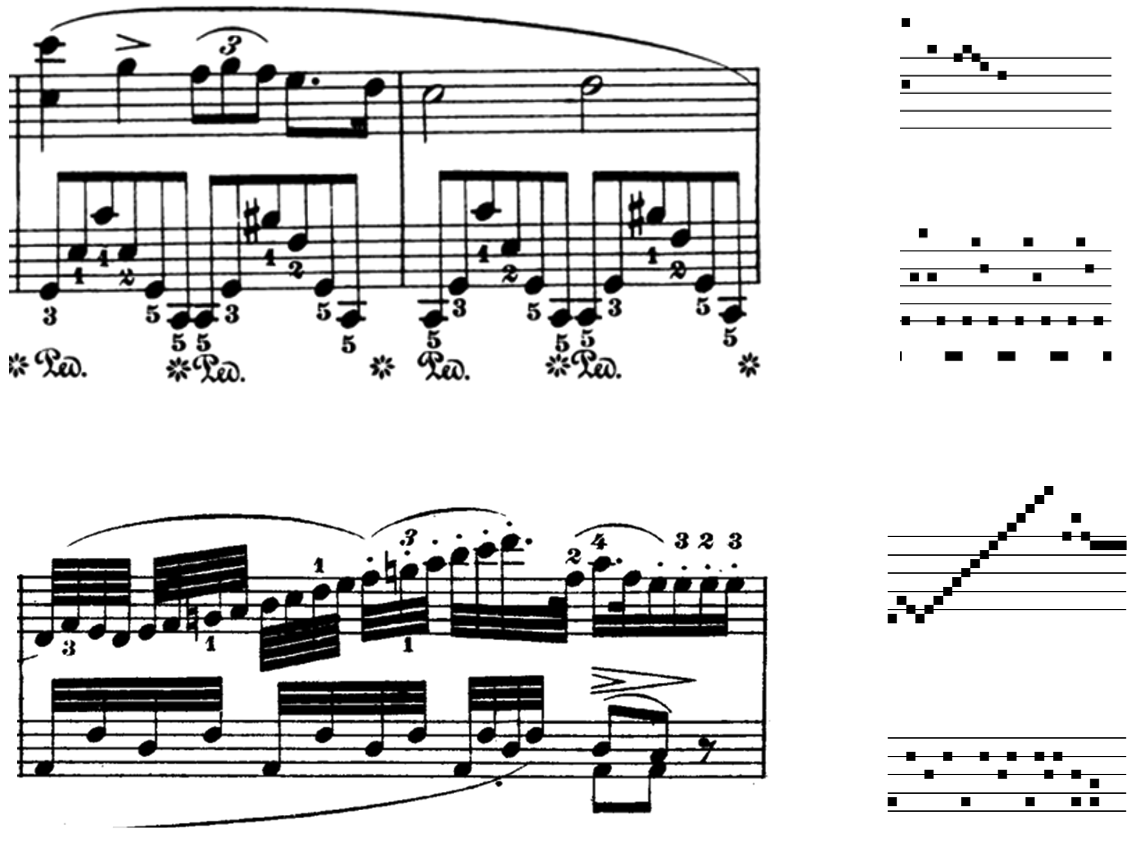}
	\caption{Two examples of a piano sheet music excerpt (left) and corresponding bootleg score representation (right).  \blue{Staff lines are not encoded in the bootleg score representation itself, but they are overlaid in the examples above as a visual reference.}}
	\label{fig:bootlegExamples}
\end{figure}

The fourth step is to compute the bootleg score representation from each PNG image (i.e., page of sheet music).  The bootleg score \citep{yang2019midi, tsai2020using} is a mid-level feature representation that encodes the position of filled noteheads relative to staff lines in the sheet music, while ignoring many other aspects of the sheet music such as note duration, accidentals, rests, time signatures, clef and octave markings, and non-filled noteheads.  \blue{The feature extraction process uses classical computer vision techniques to detect notehead and staff line locations, so it is a noisy estimation that contains errors.  More details can be found in \cite{yang2019midi}.}  Figure \ref{fig:bootlegExamples} shows two examples of a piano sheet music excerpt and its corresponding bootleg score.  \blue{Note that staff lines are not encoded in the bootleg score representation itself, but have been overlaid in Figure \ref{fig:bootlegExamples} as a visual aid.}  The bootleg score for each page of sheet music is a $62 \times L$ binary matrix, where 62 indicates the total number of different staff line positions in both the left and right hand staves and where $L$ indicates the number of detected simultaneous notehead events in the page.  \blue{We will refer to each column of the bootleg score as a bootleg score \textit{event}.  So, for example, the bottom example in Figure \ref{fig:bootlegExamples} shows a bootleg score fragment with $L=26$ events (columns), where the first bootleg score event is a 62-length array containing 60 zeros and two ``1'' entries corresponding to the noteheads at D4 and A2.}

It is worth mentioning a few practical details at this point.  First, each 62-bit bootleg score column is encoded as a single 64-bit integer, so that bootleg scores are compactly represented as a list of integers.  \blue{Each page of sheet music is thus reduced to a list of 64-bit integers, which compactly encode the bootleg score events on the page.  Since a PDF consists of multiple pages, we store the features for each page in a separate list to keep track of which page each feature comes from.}  This representation makes it possible to store an inconveniently large ($>10$TB in PNG format) dataset very compactly in memory \blue{($0.5$ GB)}.  \blue{Second, the dataset is structured as a file hierarchy separated first by composer and then by work.  This organization makes it easy to partition the data by composer or work to ensure a clean separation between different partitions.  Third,} the resulting repository after the fourth step above is the IMSLP piano bootleg scores data v1.0, which was originally presented in \cite{yang2020camera}.  \blue{The fifth and sixth steps (below) describe the improvements in the newly released v1.1 dataset.}

\begin{figure}
	\centering
	\includegraphics[width=0.5\linewidth]{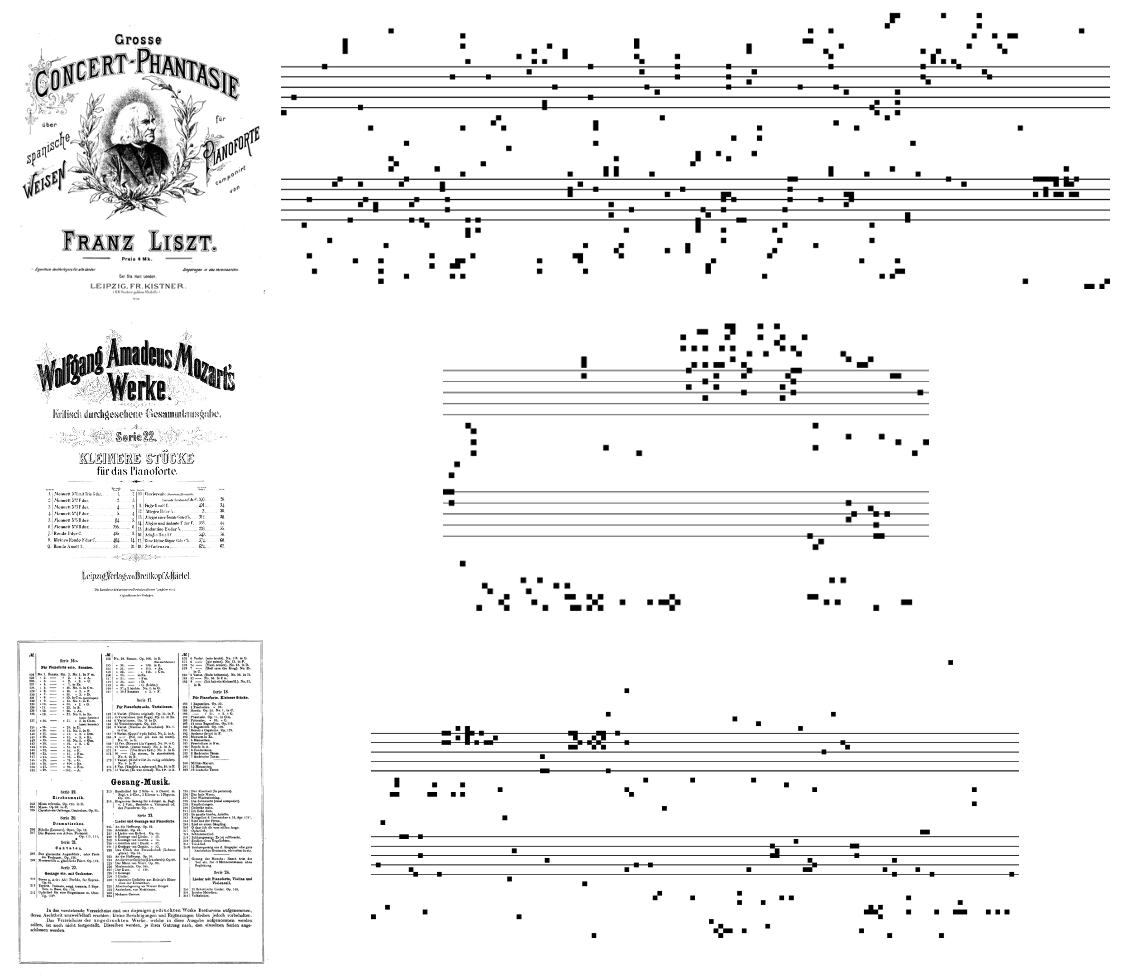}
	\caption{\blue{Examples of non-music filler pages and their extracted (gibberish) bootleg scores.}}
	\label{fig:bootlegExamplesFiller}
\end{figure}

The fifth step is to filter out non-music pages from the bootleg score repository.  One of the problems with the original repository is that many PNG images are not sheet music – they may be title pages, blank pages, foreword, table of contents, etc.  In the original v1.0 repository, a bootleg score was computed on every single PNG image, without any consideration of the contents in the image.  This results in a non-trivial amount of gibberish bootleg score data which has been extracted from non-music images.  \blue{Figure~\ref{fig:bootlegExamplesFiller} shows some examples of non-music filler pages (left) and their corresponding gibberish bootleg scores (right).}  In order to identify non-music pages, we trained a Transformer-based model to identify gibberish bootleg scores.  The process for training this model is described in detail in Section~\ref{subsec:identifyFiller}.  In the revised v1.1 IMSLP piano bootleg scores repository, the bootleg scores for (predicted) non-music pages have been identified and removed.

\blue{The sixth step is to scrape, organize, and include metadata that is available on IMSLP for each work in the unlabeled dataset.  This process is described in Section~\ref{subsec:imslpMetadata}.  This metadata is included in the v1.1 IMSLP piano bootleg scores repository to facilitate multimodal research and to allow for convenient linking to other datasets.}

\subsection{Identifying Non-music Pages}
\label{subsec:identifyFiller}

In this subsection, we describe the process of identifying non-music pages by training a Transformer-based model on bootleg score fragments.

The first step is to label a set of music and non-music pages.  This was done in the following manner.  First, we took the original 9-class dataset proposed in \cite{tsai2020composer}, in which each page had been manually labeled as music or non-music.  We manually re-labeled these pages into three categories: music, non-music, or mixture.  The “mixture” category contains pages that contain both sheet music and text, as is often seen in a table of contents (e.g.,~showing excerpts of pieces) or foreword.  We ultimately decided to exclude the mixture pages from training, and only include pure music and pure non-music pages for training our classifier.  In total, there were 5938 music pages and 259 non-music pages.  We divided these pages into training and validation partitions using a 60-40 split. 

\begin{figure}
	\centering
	\begin{subfigure}[b]{0.5\linewidth}
		\includegraphics[width=1\linewidth]{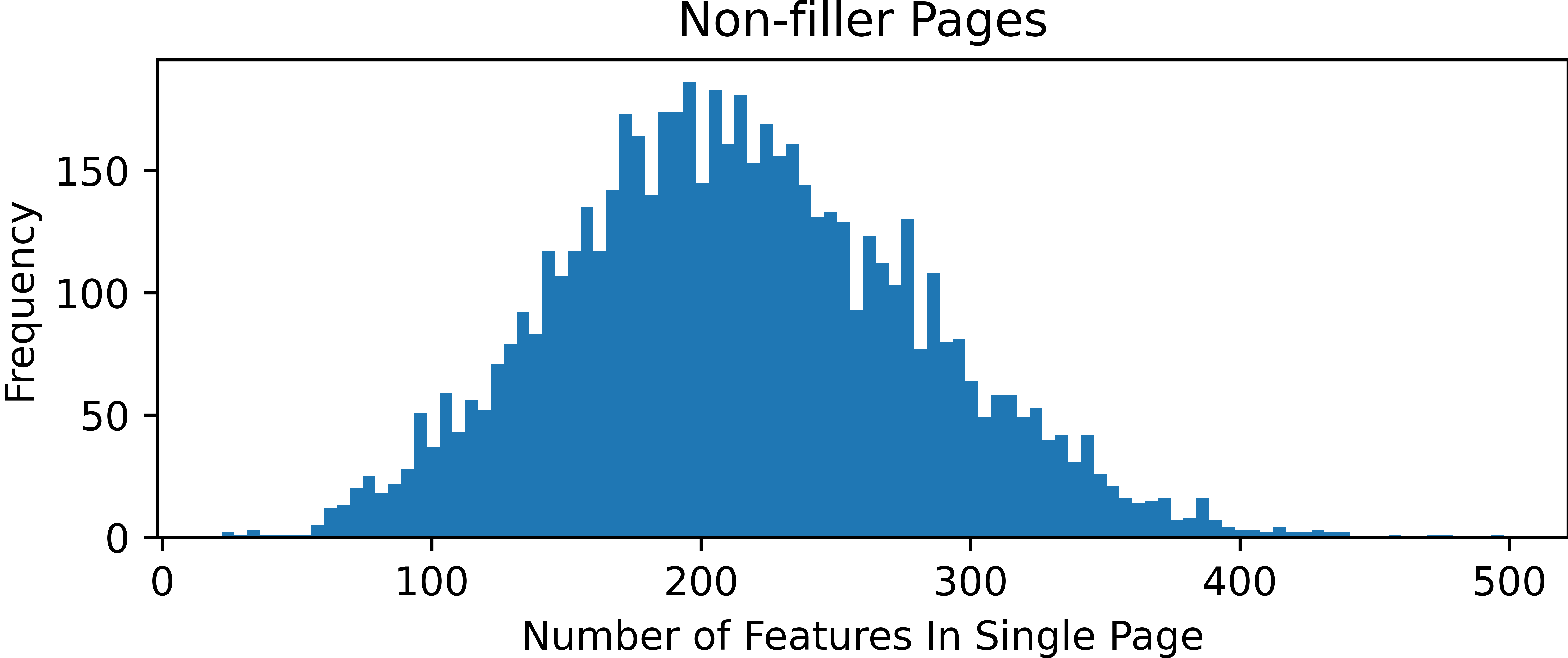}
	\end{subfigure}\vspace{5mm}
	\begin{subfigure}[b]{0.5\linewidth}
		\includegraphics[width=1\linewidth]{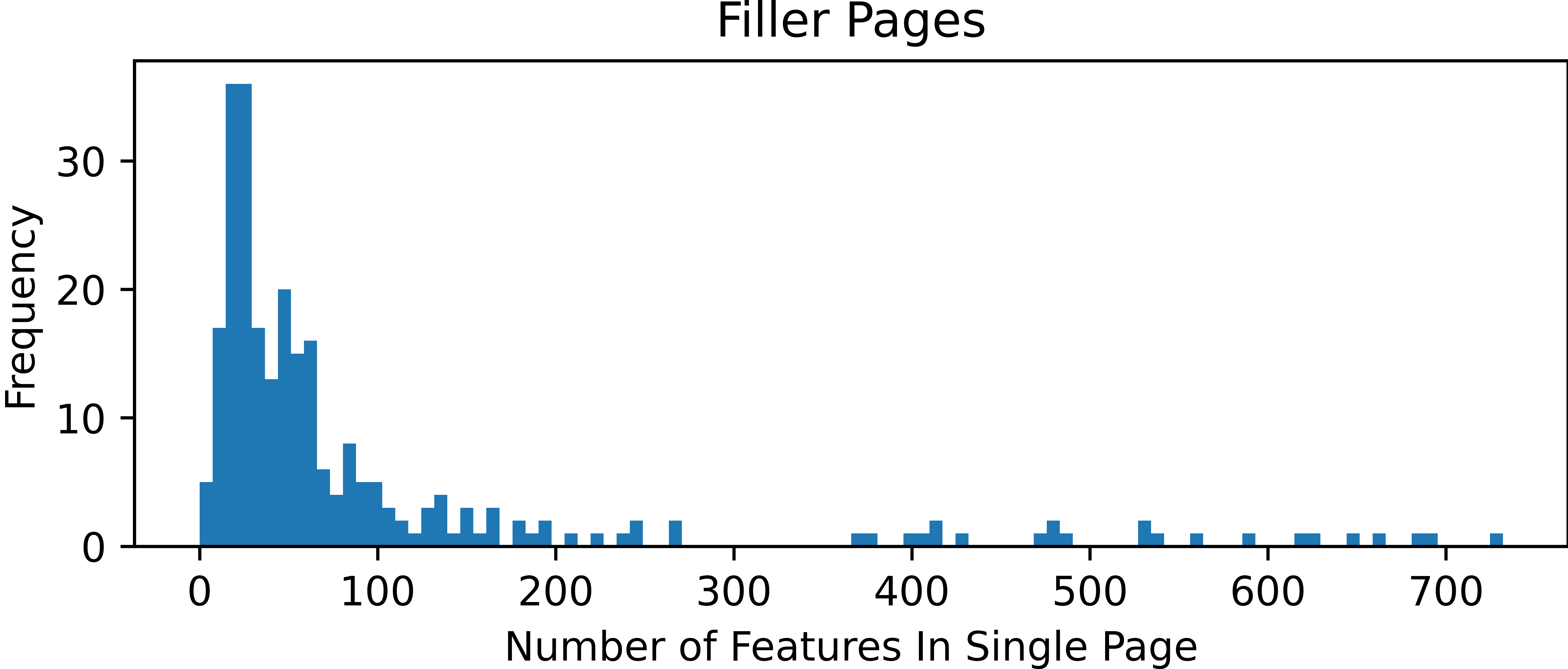}
	\end{subfigure}
	\caption{Histogram of the number of bootleg score events in a set of manually labeled music pages (top) and non-music pages (bottom).}
	\label{fig:fillerHistograms}
\end{figure}

The second step is to sample bootleg score fragments.  Figure \ref{fig:fillerHistograms} shows a histogram of the number of bootleg score events in music pages (top) and non-music pages (bottom).  We can see that many filler (non-music) pages have a very small number of bootleg score events, so our classifier will need to handle short bootleg score fragments.  Accordingly, we decided to train our model on bootleg score fragments of length 16.  We densely sampled bootleg score fragments from the non-music pages by sampling 16-length fragments with 50\% overlap.  This resulted in a total of 2799 non-music bootleg score fragments (1689 train, 1110 validation).  To maintain a balanced dataset, we randomly sampled the same number of fragments from the music pages.  This sampling was done by randomly sampling a work (PDF) from the train/validation partition, randomly sampling a music page from the PDF, and then randomly sampling a length 16 fragment from the page’s bootleg score.  At the end of this step, we have a labeled dataset of 3378 training bootleg score fragments (1689 filler, 1689 non-filler) and 2220 validation bootleg score fragments (1110 filler, 1110 non-filler).

The third step is to train a music vs non-music fragment classifier.  We adopted a similar approach as in \cite{tsai2020composer}, which we describe here for completeness.  We encode each 62-bit bootleg score column as a sequence of eight 8-bit characters, and learn a subword vocabulary using Byte Pair Encoding \citep{gage1994new}.  Using this BPE tokenizer, we pretrain a GPT-2 language model on the entirety of the IMSLP piano bootleg scores repository (v1.0).  \blue{It is worth noting that this pretrained language model was originally used for composer classification, and here we simply finetuned it on a different downstream task.}  Next, we add a classification head with two output classes (music vs non-music) and finetune it on the labeled dataset of music and non-music fragments.  In this way, our classifier is trained to classify 16-length bootleg score fragments as music or non-music.

We apply our classifier model to full pages in the following manner.  We first extract a bootleg score representation from the page.  If the resulting bootleg score has a length less than 64, it is automatically classified as non-music.  (Note from Figure \ref{fig:fillerHistograms} that very few music pages have bootleg score lengths less than 64.)  Otherwise, fragments of length 16 are densely sampled from the bootleg score with 50\% overlap, and each fragment is passed through our classifier model.  We average the outputs of each fragment prediction to get an ensembled prediction for the entire page.

\begin{figure}
	\centering
	\includegraphics[width=0.5\linewidth]{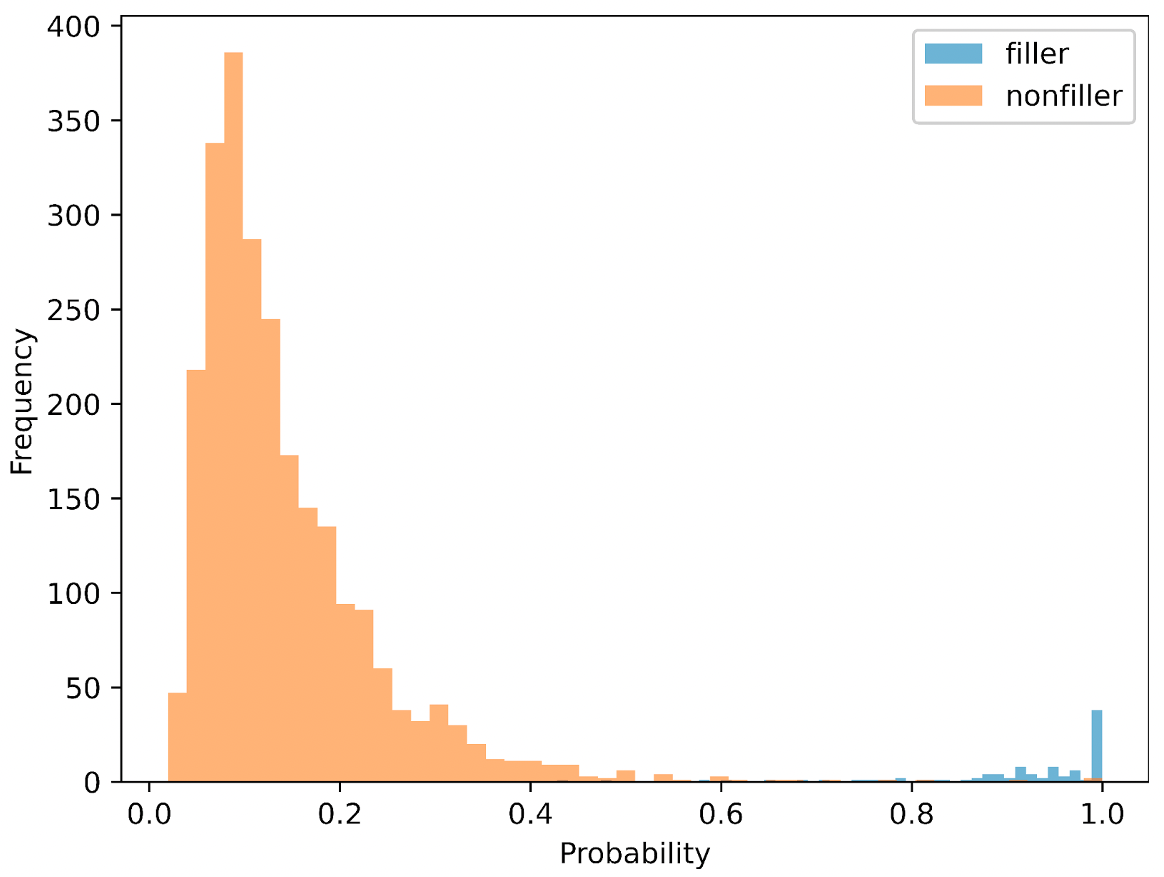}
	\caption{Predicted probability of an ensembled classifier that classifies validation pages as filler (non-music) vs non-filler.  We use a hard threshold of 0.5 to ensure that filler pages are excluded from our dataset with high confidence.}
	\label{fig:fillerModelDistr}
\end{figure}

Figure \ref{fig:fillerModelDistr} shows a histogram of predicted probabilities on the validation pages, where a higher probability corresponds to a non-music page.  We can see that there is a fairly clean separation between the music and non-music data.  We set a very conservative threshold of 0.5, which ensures that non-music pages will be excluded from the data with high confidence (and sometimes music data will be excluded as well, which we are okay with).  With this threshold value, we achieve a precision of 0.85 and a recall of 0.98 on the validation pages.  Because we care more about ensuring that non-music pages are excluded, the recall of 0.98 is the more important metric.  We use this ensembled classifier to identify and remove non-music pages from the IMSLP piano bootleg score repository.

\subsection{Adding Metadata from IMSLP}
\label{subsec:imslpMetadata}

\blue{In addition to cleaning up the IMSLP piano bootleg scores repository, we also collected and added metadata for these works.  This process is described below.}

\blue{The metadata is scraped from the IMSLP website.  IMSLP has a webpage for each work which contains links to audio performances and various sheet music editions.  The webpage also contains a set of metadata for the composition, which may include attributes such as the work title, composer, opus/catalogue number, key, year/date of composition, composer time period, instrumentation, movements/sections, alternative title, dedication, first publication, etc.  We scraped the composition webpages to extract these metadata attributes, and have stored the metadata in a file on our github repository.  }

\blue{This metadata is valuable for two reasons.  First, it provides a much richer set of information that could be used to study many other tasks besides composer classification.  Second, it allows for convenient linking to other datasets.  As one concrete example, we used simple string matching based on the composer name and work title attributes to link the bootleg scores in the PBSCR dataset with corresponding files in the GiantMIDI-Piano dataset \citep{kong2020large}.  A file containing 7413 matches is included in our github repository.  The provided metadata can similarly be used to link the PBSCR dataset to other datasets.}

\section{Dataset Preparation: Labeled Data}
\label{sec:dataLabeled}

\blue{In this section, we describe the preparation of the labeled (100-class, 9-class) PBSCR data.  The 100-class (Section~\ref{subsec:data100way}) and 9-class (Section~\ref{subsec:data9way}) data provide labeled bootleg score fragments to train and evaluate models for the composer recognition task.  These datasets have been designed to make the data as accessible and easy to use as MNIST, in order to enable rapid iteration and experimentation.  Compared to the labeled dataset used in \cite{tsai2020composer}, the novel contributions are to expand the number of composer classes from 9 to 100 (Section~\ref{subsec:data100way}) and to offer a discussion of known data leakage issues (Section~\ref{subsec:dataLeakage}).  The 9-class labeled dataset is included to enable experimentation on tasks of varying difficulty and to allow for historical comparisons to previous work.}

\subsection{100-class Labeled Data}
\label{subsec:data100way}

\blue{The preparation of the 100-class labeled dataset consists of four steps, which are described below.  At a high level, it consists of $100,000$ $62\times64$ bootleg score fragments (70k train, 15k validation, 15k test) that are balanced across 100 different classical composers.}

The first step is to identify a list of 100 composers to include.  \blue{These 100 composers are selected as a subset from the IMSLP Piano Bootleg Scores Data described in Section~\ref{subsec:imslpData}.}  We first ranked all composers by the total amount of bootleg score data they have available on IMSLP.  We then manually reviewed the ordered list of composers and selected the top 100, being sure to remove those who are not primarily composers (e.g., some people on the list were primarily arrangers and editors).  \blue{The full list of 100 composers can be found at \url{https://github.com/HMC-MIR/PBSCR/blob/main/100_class_list.txt}.}

\begin{figure}
	\centering
	\includegraphics[width=0.5\linewidth]{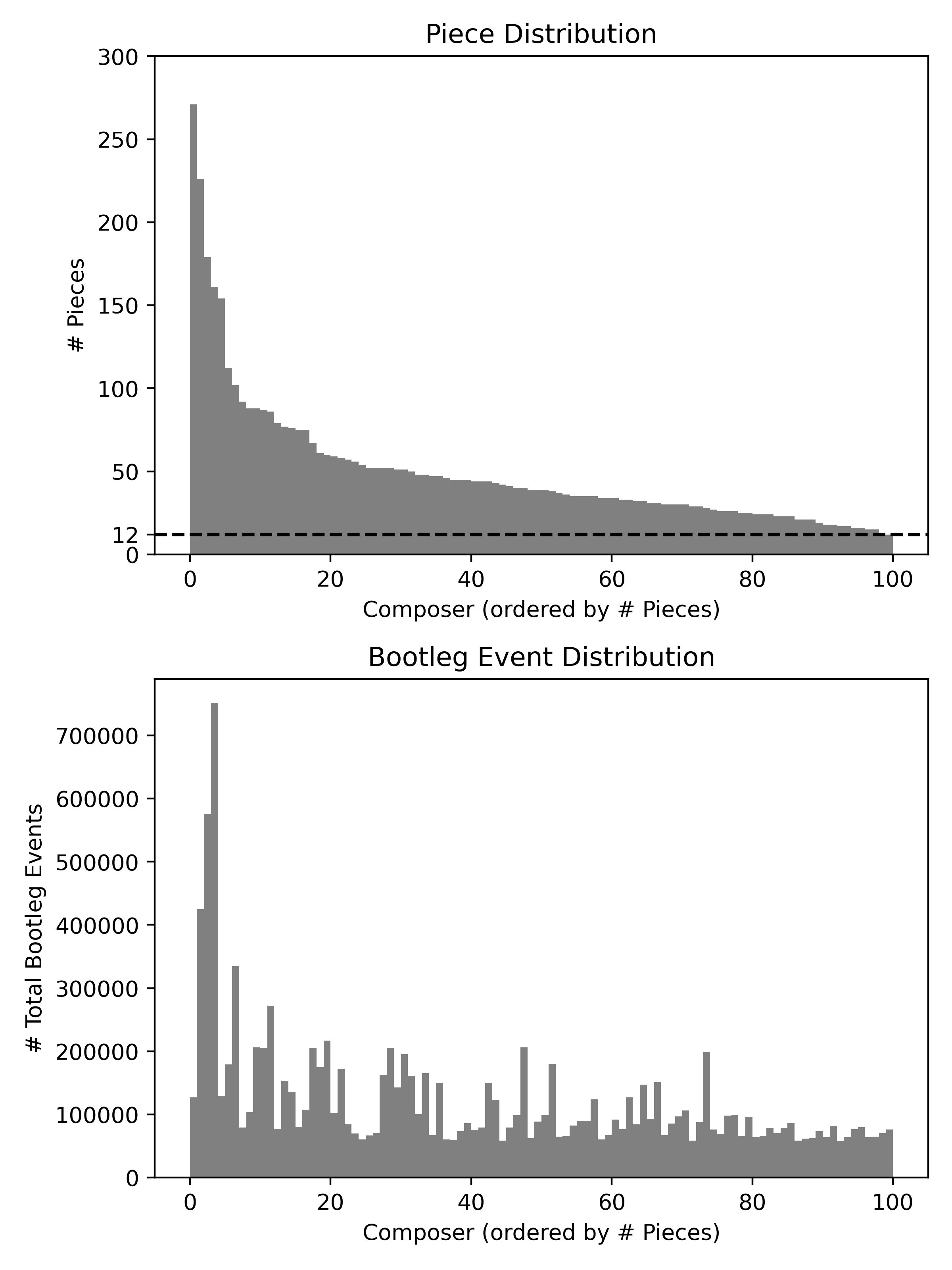}
	\caption{(Top) The total number of pieces/works available on IMSLP for the composers in the 100-class dataset.  (Bottom) The total number of bootleg score events for each composer in the 100-class dataset.  \blue{The list of composers sorted by number of works can be found at \url{https://github.com/HMC-MIR/PBSCR/blob/main/forPaper/composers_sorted_numpieces.txt}.}}
	\label{fig:dataHistograms}
\end{figure}

The second step is to select a set of sheet music PDFs for each composer.  Each work in IMSLP may have multiple PDFs associated with it, which correspond to different publishers or editions.  Because popular works tend to have a large number of sheet music versions, we select one representative PDF per work in order to avoid over-representing a small number of works.  In order to maximize the amount of data available to us, we simply selected the PDF that had the highest number of total bootleg score events.  Figure \ref{fig:dataHistograms} shows the total number of works available for these top 100 composers (top), \blue{along with the total number of bootleg score events extracted from each composer's piano sheet music (bottom).}  \blue{In total, there are 4997 works and 70440 sheet music pages.}

The third step is to identify non-music pages in the selected set of PDFs.  We used a Transformer-based model to identify filler pages, as described in Section~\ref{subsec:identifyFiller}.  \blue{In the 100-class labeled data, there are a predicted 64129 (out of 70440) pages with music content with $12.1$ million bootleg score events.}

The fourth step is to sample bootleg score fragments from each composer.  \blue{This sampling serves two purposes: it allows us to achieve class balance among the fragments even though the number of works per composer is different, and it standardizes the size of each labeled sample ($62\times64$) in order to achieve the data simplicity of MNIST images.  This sampling is done in the following manner.}  First, we divide the works into training, validation, and test sets, using a split of 70\%, 15\%, and 15\%, respectively.  Next, we decided on the total number of bootleg score fragments to sample from each partition.  For the 100-class data, we have 70000 train fragments, 15000 validation fragments, and 15000 test fragments, resulting in a total of 100,000 examples.  \blue{Based on these numbers, we calculated how many fragments per composer need to be sampled in order to achieve class balance.  Each fragment is drawn by randomly selecting a work by a given composer, and then randomly selecting a 64-length fragment from the bootleg score.  Our sampling process guarantees that our classes are perfectly balanced, and it gives equal weight to all piano works (in IMSLP) that a composer has composed.}

\begin{figure}
	\centering
	\includegraphics[width=0.5\linewidth]{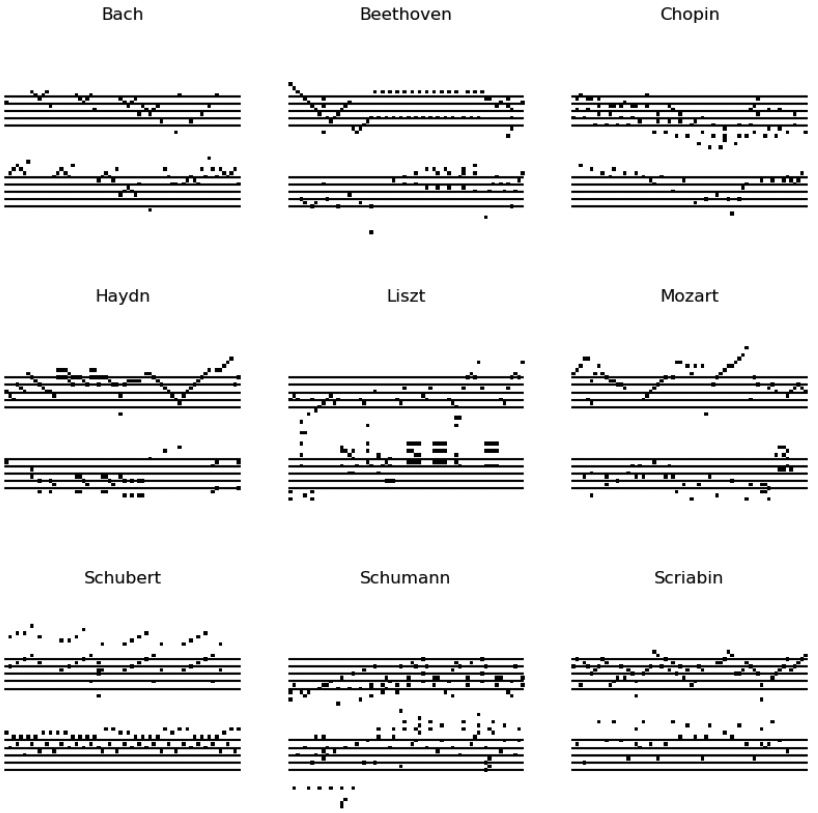}
	\caption{Example bootleg score images from the labeled 9-class PBSCR data.  \blue{Staff lines have been overlaid for ease of interpretation.}}
	\label{fig:labeledExamples}
\end{figure}

Figure \ref{fig:labeledExamples} shows a set of example bootleg score images for \blue{9 selected composers (those in the 9-class dataset).  The staff lines in the left and right hands are not present in the bootleg score representation itself, but they have been overlaid for ease of reference.}  Even without any information about note durations, key or time signature, or accidentals, one can immediately see some recognizable features: the Bach example has fugue-like texture and movement.  The Beethoven example has an alternating octave in the right hand, which is not common in Bach's music.  The Mozart example has scale-like runs in the right hand with an Alberti bass-like left hand accompaniment.  The Classical and Baroque composers (Bach, Mozart, Haydn) have thinner textures compared to the Romantic era composers.  These examples show that, even with the minimal bootleg score representation, many aspects of compositional style are preserved.

\blue{The 100-class labeled dataset is formatted in a way that resembles the MNIST dataset.  Each dataset consists of the following six arrays:}
\begin{itemize}
	\item \blue{$X_{\mathrm{train}}$: a $70000 \times 62 \times 64$ binary tensor specifying the training bootleg score fragments}
	\item \blue{$Y_{\mathrm{train}}$: a $70000$ length array specifying the train composer class indices}
	\item \blue{$X_{\mathrm{valid}}$: a $15000 \times 62 \times 64$ binary tensor specifying the validation bootleg score fragments}
	\item \blue{$Y_{\mathrm{valid}}$: a $15000$ length array specifying the validation composer class indices}
	\item \blue{$X_{\mathrm{test}}$: a $15000 \times 62 \times 64$ binary tensor specifying the test bootleg score fragments}
	\item \blue{$Y_{\mathrm{test}}$: a $15000$ length array specifying the test composer class indices}
\end{itemize}
In addition, we also provide relevant metadata on all train, validation, and test fragments.  This metadata includes a unique identifier that specifies the PDF from IMSLP from which the fragment was taken, as well as the page and offset in the bootleg score from which the fragment was sampled.  \blue{This information allows researchers to access the complete, unabridged bootleg scores to study the effect of longer-term structure, or to access the original raw sheet music image data for visualization and deeper understanding.}

\blue{The 100-class dataset poses a much more challenging classification task than the 9-class recognition task in \cite{tsai2020composer}}.  One may notice that the shape and size of the data is similar to MNIST, in keeping with our motto of constructing a dataset that is ``as accessible as MNIST and as challenging as ImageNet.''  Each composer only has 700 training examples, so the task is difficult both for the large number of composers and the relative scarcity of labeled data.  For these reasons, we believe this dataset will push the boundaries of composer recognition to the next level.

\subsection{9-class Labeled Data}
\label{subsec:data9way}

\blue{The 9-class labeled dataset contains bootleg score fragments for 9 classical composers: Bach, Beethoven, Chopin, Haydn, Liszt, Mozart, Schubert, Schumann, and Scriabin.  The dataset is constructed in the same way as the 100-class labeled dataset but with one difference: the list of composers was adopted from \cite{tsai2020composer} (rather than selected based on data availability) to enable historical comparisons.  In total, the 9-class dataset consists of 28000 train fragments, 6000 validation fragments, and 6000 test fragments, resulting in a total of 40000 examples that are balanced across composers.  Table \ref{tab:datasetSummary} shows the number of works, total number of pages, number of (predicted) music pages, and number of bootleg score events for each composer in the 9-class labeled dataset.  There are 896 PDFs, 10305 pages with music content, and $2.2$ million bootleg score events.  The purpose of providing both 100-class and 9-class datasets is to enable experimentation at varying levels of task difficulty.}

\begin{table}
	\centering
	\begin{tabular}{lccc}
		\toprule
		\bfseries Composer & \# Works & \# Pages & \# Bootleg \\
		\bfseries & & (all/music) & Features \\  \midrule
		Bach & 226 & 1752/1666 & 424948 \\
		Beethoven & 86 & 1292/1170 & 272374 \\
		Chopin & 89 & 1048/996 & 205513 \\
		Haydn & 51 & 50/50 & 12408 \\
		Liszt & 179 & 3405/3170 & 575367 \\
		Mozart & 61 & 702/673 & 174355 \\
		Schubert & 88 & 836/836 & 206103 \\
		Schumann & 40 & 981/919 & 206379 \\
		Scriabin & 76 & 879/825 & 135851 \\ \midrule
		9-class & 896 & 10945/10305 & 2213298 \\
		100-class & 4997 & 70440/64129 & 12108749 \\
		\bottomrule
	\end{tabular}
	\caption{Overview of the raw sheet music data from which the 9-class PBSCR data was constructed.  Cumulative counts for the 100-class data are also shown at the bottom.}
	\label{tab:datasetSummary}
\end{table}

\subsection{\blue{Data Leakage}}
\label{subsec:dataLeakage}

\blue{In this section we discuss known data leakage issues with the 9-class and 100-class labeled datasets.}

\blue{It is impossible to get a perfectly clean train/test split with IMSLP data for many reasons: composers often re-write pieces or re-use themes in later works (and each may be listed as a separate opus number), composers sometimes create alternate versions of pieces, some composers do transcriptions or arrangements of other composers' works (which makes the ground truth label ambiguous), some works are partially composed by the composer and completed by others after the composer's death, and some works have uncertain or incorrect authorship (e.g.,~the Valse melancolique in F-sharp minor was incorrectly attributed to Chopin and listed among his list of works).}

\blue{We discovered an additional source of data leakage late in the review process: idiosyncrasies in IMSLP's organization.  Specifically, there are instances where the same piece may appear on two different IMSLP webpages: once as an individual composition and once as a part of a collection (e.g.,~individual preludes and fugues in Bach's \textit{Well Tempered Clavier}).  To quantify how often this happens, we manually checked all works in the 9-class dataset and found that 11 of the 896 works were collections exhibiting this issue (9 Bach, 1 Scriabin, 1 Liszt).  Of these 11 collections, we noticed that the IMSLP webpages for eight of them had a warning dialog box at the top of the webpage that indicated the possibility of duplicate entries.  The 100-class dataset was too large to check manually, but we used an automated approach to iterate through all of the 4997 works and detect the warning dialog box mentioned above.  We found only two more collections with this issue (1 Mendelssohn, 1 Handel).  Thus, we found 13 works/collections across 5 composers in the 100-way dataset that exhibit this issue.  This automated approach most likely failed to catch some additional instances, but it provides a ballpark estimate of how common this phenomenon is.  A list of these collections can be found on our github repository.}

\blue{The above sources of data leakage mean that our reported accuracy numbers are likely inflated.  However, as long as researchers use the same dataset for comparison, the benchmark can still serve its purpose to track progress on the composer classification task.  Also, given how low the accuracies are for the 100-way classification (e.g., the best top-1 accuracy is $13.9\%$ in Table \ref{tab:results100way}), the accuracy inflation due to train/test leakage still leaves an enormous amount of room for improvement and is likely only a minor factor in overall performance.}

\section{Research Tasks}
\label{sec:researchTasks}

In this section, we describe several research tasks that could be studied with the PBSCR dataset.  We also provide baseline results using standard techniques for future researchers to compare against.

\subsection{\blue{Supervised} composer recognition}
\label{subsec:tasks_MBCSR}

The most obvious task is a \blue{supervised} composer recognition task.  Here, the goal is to classify a bootleg score fragment according to its composer class.  In addition to a \blue{labeled set of training pairs, unlabeled data from the IMSLP Piano Bootleg Scores v1.1 dataset (Section~\ref{subsec:imslpData})} is available for pretraining.

There are several metrics of performance that might be appropriate in this scenario.  Following the convention in ImageNet, one useful metric of performance is top $N$ accuracy, which indicates the percentage of queries that have the correct composer in the $N$ highest-ranked composers.  Another useful metric is mean reciprocal rank (MRR), which is calculated as 
\begin{equation}
	MRR = \frac{1}{N} \sum_{i=0}^{N-1} \frac{1}{R_i}
\end{equation}
where $R_i$ indicates the rank of the true composer.  \blue{MRR ranges between $0$ and $1$, where a higher MRR is better.  MRR offers more nuanced information about the rank of the true composer than the hard binary threshold of a top $N$ accuracy metric.}  We recommend reporting results with several of the above metrics, since the most appropriate metric may depend on the difficulty of the task.

Tables \ref{tab:results9way} and \ref{tab:results100way} show the performance of three different baseline systems on the 9-class and 100-class recognition tasks, respectively.  The first baseline system is a CNN model with two convolutional layers, followed by global average pooling across time, and then a final output linear classification layer.  This model is based on the architecture proposed in \citep{verma2019midi} but adapted to a bootleg score representation (instead of MIDI).  The second baseline system is a GPT-2 model \citep{radford2019language} trained in the same manner as in Section~\ref{subsec:identifyFiller}: each bootleg score column is represented as a sequence of 8-bit characters, a subword vocabulary is learned using Byte Pair Encoding, a small 6-layer GPT-2 language model is trained on unlabeled bootleg scores in IMSLP, and the model is fine-tuned on the labeled data.  To tease apart the effect of pretraining and finetuning, we report results of the GPT-2 model under three different training conditions: (1) training the model from scratch on the labeled data without pretraining (``GPT-2 (no pretrain)''), (2) pretraining the language model and learning a linear probe (``GPT-2 (LP)''), and (3) pretraining the language model, learning a linear probe, and then unfreezing and finetuning the whole model (``GPT-2 (LP-FT)'').  For (3), we followed the recommended practices in \cite{kumar2022fine}, which were shown to have good generalization to out-of-distribution data.  The third baseline system is a RoBERTa model \citep{liu2019roberta} with 6 Transformer encoder layers.  This model is pretrained using a masked language modeling task, but otherwise trained in a similar manner as GPT-2.  We report results of the RoBERTa model under the same three training conditions as above.  These three model architectures were previously explored in \cite{tsai2020composer} and \cite{yang2021composer} on a 9-class composer recognition task, and here we present results on the (new) 9-class and 100-class PBSCR benchmarks.  These results are intended to serve as baselines which future approaches can compare against.

\begin{table}
	\centering
	\begin{tabular}{lcc}
		\toprule
		\bfseries System & Top 1 & MRR\\ \midrule
		CNN & 40.0 & 0.593 \\ 
		GPT-2 (LP-FT) & 49.6 & 0.670\\
		GPT-2 (LP) & 42.5 & 0.613\\
		GPT-2 (no pretrain) & 25.0 & 0.466\\ 
		RoBERTa (LP-FT) & 44.4 & 0.631 \\
		RoBERTa (LP) & 38.0 & 0.581 \\
		RoBERTa (no pretrain) & 19.2 & 0.407 \\
		\bottomrule
	\end{tabular}
	\caption{Baseline results for the 9-class PBSCR task.  Results are shown for top 1 accuracy (\%) and mean reciprocal rank.}
	\label{tab:results9way}
\end{table}

\begin{table}
	\centering
	\begin{tabular}{lccc}
		\toprule
		\bfseries System & Top 1 & Top 5 & Top 10\\ \midrule
		CNN & 7.4 & 21.3 & 32.4 \\
		GPT-2 (LP-FT) & 13.9 & 34.8 & 49.0 \\
		GPT-2 (LP) & 10.4 & 28.5 & 42.8 \\
		GPT-2 (no pretrain) & 3.2 & 11.6 & 20.4 \\
		RoBERTa (LP-FT) & 10.6 & 29.0 & 42.0 \\
		RoBERTa (LP) & 7.5 & 22.9 & 35.0 \\
		RoBERTa (no pretrain) & 2.1 & 8.1 & 15.0 \\
		\bottomrule
	\end{tabular}
	\caption{Baseline results for the 100-class PBSCR task.  Results are shown for top 1, top 5, and top 10 accuracy (\%).}
	\label{tab:results100way}
\end{table}

There are two things to notice about the baseline results in Tables \ref{tab:results9way} and \ref{tab:results100way}.  First, the GPT-2 model has the best performance among the three models on both the 9-class and 100-class recognition tasks.  We can see that pretraining on the unlabeled IMSLP data makes a big difference, improving top-1 accuracy on the 9-class recognition task from 25.0\% to 42.5\% and improving top-5 accuracy on the 100-class recognition task from 11.6\% to 28.5\%.  This underscores the importance of having a large, diverse set of data for pretraining.  We also see that full model fine-tuning makes a big difference, improving top-1 accuracy on the 9-class recognition task from 42.5\% to 49.6\% and improving top-5 accuracy on the 100-class recognition task from 28.5\% to 34.8\%.  Second, there is a lot of room for improvement.  The best GPT-2 model only achieves a top-5 accuracy of 34.8\% on the 100-class recognition task, showing that there is a massive amount of room for improvement.  Our hope is that this dataset can spur progress on this challenging task.

\subsection{\blue{1-shot and low-shot} composer recognition}
\label{subsec:data_fewshot}

An interesting modification to the above problems is to study few-shot composer recognition.  The problem setup would be the same as before, but the number of training examples per composer would be artificially limited to $N$.  We recommend the following tasks: (a) a 9-class recognition task with $N=1$, $10$, $100$ and (b) a 100-class recognition task with $N=1$, $10$, $100$.  This set of tasks encourages the development of approaches that are data-efficient (with labeled data), and it allows one to study the effect of the number of training examples as well as the generalizability of model representations.

\begin{table}
	\centering
	\begin{tabular}{lccccc}
		\toprule
		\bfseries System & N & Top 1 & Top 1 & MRR & MRR \\
		\bfseries &  & mean & std &  mean & std\\ \midrule
		GPT-2 & 1 & 15.4 & 2.3 & 0.36 & .020 \\
		RoBERTa & 1 & 14.5 & 1.8 & 0.35 & .017 \\ 
		Random & 1 & 11.2 & 0.3 & 0.32 & .003 \\ \midrule
		GPT-2 & 10 & 19.7 & 1.8 & \blue{0.41} & .013 \\
		RoBERTa & 10 & 19.8 & 1.6 & \blue{0.41} & .013 \\ 
		Random & 10 & 11.0 & 0.4 & \blue{0.31} & .003 \\ \midrule
		GPT-2 & 100 & 23.8 & 0.8 & \blue{0.45} & .006 \\
		RoBERTa & 100 & 23.7 & 0.9 & \blue{0.45} & .006 \\
		Random & 100 & 11.1 & 0.4 & \blue{0.31} & .004 \\
		\bottomrule
	\end{tabular}
	\caption{Baseline results for the $N$-shot 9-class recognition task.  Results are expressed as a mean and standard deviation across 30 trials.  Top 1 accuracies are indicated in percentages (\%).}
	\label{tab:fewshotResults9way}
\end{table}

\begin{table*}
	\centering
	\begin{tabular}{lccccccc}
		\toprule
		\bfseries System & N & Top 1 & Top 1 & Top 5 & Top 5 & Top 10 & Top 10 \\
		\bfseries &  & mean & std &  mean & std & mean & std\\ \midrule
		GPT-2 & 1 & 1.9 & .21 & 7.7 & .44 & 14.1 & .56 \\
		RoBERTa & 1 & 1.8 & .20 & 7.7 & .45 & 14.1 & .57\\
		Random & 1 & 1.0 & .06 & 5.0 & .13 & 10.0 & .21\\ \midrule
		GPT-2 & 10 & 3.0 & .25 & 11.2 & .38 & 19.1 & .50\\
		RoBERTa & 10 & 3.1 & .19 & 11.3 & .39 & 19.3 & .54\\
		Random & 10 & 1.0 & .10 & 5.0 & .15 & 10.0 & .23\\ \midrule
		GPT-2 & 100 & 3.9 & .17 & 14.2 & .30 & 23.5 & .41\\
		RoBERTa & 100 & 4.0 & .14 & 14.3 & .27 & 23.7 & .34\\
		Random & 100 & 1.0 & .07 & 5.0 & .16 & 10.0 & .21\\
		\bottomrule
	\end{tabular}
	\caption{Baseline results for the few-shot 100-class recognition task.  Results are expressed as a mean and standard deviation of test set accuracy (\%) across 30 trials.}
	\label{tab:fewshotResults100way}
\end{table*}

We consider three baseline models for the few-shot tasks.  The first system is a GPT-2 model that is trained by: (a) pretraining a  6-layer GPT-2 language model on the unlabeled IMSLP bootleg score data, as described in Section \ref{subsec:tasks_MBCSR}, (b) using the penultimate activations of the language model as a feature representation for the training samples, (c) identifying the $k$ nearest neighbors for each composer that are closest in euclidean distance to a given test query, and (d) rank ordering the composers by average euclidean $k$ nearest neighbor distance.  The second system is a 6-layer RoBERTa model that is trained and used in a similar manner, but using a masked language modeling task during pretraining.  These models adopt the pretraining strategies of the classification models described in Section \ref{subsec:tasks_MBCSR}, but use $k$ nearest neighbors for classification instead of training a classification layer.  We also evaluate the performance of a random guessing baseline for reference.

Tables \ref{tab:fewshotResults9way} and \ref{tab:fewshotResults100way} show the performance of the three baseline systems on the few-shot 9-class and 100-class recognition tasks, respectively.  The upper, middle, and bottom sections of the table show performance for an $N$-shot task with $N=1$, $N=10$, and $N=100$, respectively.  For $N=1$ we use the $k=1$ nearest neighbor for each composer (by necessity), and for $N=10$ and $N=100$ we use the $k=3$ nearest neighbors.  In each trial, we randomly sample $N$ training examples from each composer to simulate a few-shot scenario, and then calculate the performance on the entire test set.  We report the mean and standard deviation of performance across 30 trials.

We can see that the pretrained models perform significantly better than random, and that both models perform comparably across all settings.  While these results show that the pretrained models are indeed extracting style information, the performance of these models is quite poor overall, indicating how much room there is for improvement.  We provide these results as a baseline against which future works can compare.

\subsection{0-shot composer recognition}
\label{subsec:data_0shot}

Another interesting problem to consider is zero-shot composer recognition.  In this task, the goal is to predict the composer of a bootleg score fragment when no previous training examples of that composer have been seen.  We are not aware of previous work studying this topic within the composer recognition literature until very recently, where researchers explore zero-shot composer classification with music-text data \citep{wu2023clamp}.  Here, we simply define the task and suggest some possible avenues of exploration.

This task is possible due to the rich metadata available on IMSLP.  For example, given the Wikipedia articles for some unknown composers, one could infer aspects of compositional style based on their date of birth, country of origin, connections to other composers, or other knowledge about the composer that is embedded in a large language model.  \cite{wu2023clamp} train a model to embed both symbolic music and text descriptions into a common embedding space using a CLIP-like approach \citep{radford2021learning}.  This work does not release their music-text training pairs, however, and also evaluates on a small evaluation dataset (411 pieces, 8 classes).  The PBSCR data would be sufficiently large to train such models, is open to the research community, and offers a much more challenging classification task.  Setting up a benchmark for multimodal tasks such as this is an area for future work.

The zero-shot task opens the door to multimodal approaches to composer recognition.  Given the size, richness of metadata, and open nature of IMSLP, we believe that the PBSCR data is well poised to facilitate interesting and novel directions in multimodal research.

\section{Research Questions}
\label{sec:researchQuestions}

In this section, we describe several research questions that we believe the PBSCR data is especially well suited to facilitate research on.  

\subsection{Encoding Schemes}
\label{subsec:questions_encoding}

One open research question is, ``How should we encode music data when feeding it into a model?"  We may want to select the encoding scheme to maximize performance on a particular task of interest, to have certain desirable properties such as key or tempo invariance, or some combination of factors.  Because of our design decision to use a bootleg score representation, the PBSCR data has discarded a significant amount of musical information.  Nonetheless, due to its simple format, it is well poised to facilitate rapid, iterative exploration of many interesting questions, some of which we describe below.

\textit{Image vs Tokens}.  The fact that the bootleg score is a binary matrix raises the question: Is it better to treat the data as a 2D binary image or as a sequence of discrete tokens (e.g.,~each bootleg score column is interpreted as a discrete ``word")?  These two options lead to different kinds of models: 2D images lend themselves to CNN or ViT-based architectures, while token sequences lend themselves to Transformer-based language models.  Previous work \citep{yang2021composer} has compared simple CNN architectures with GPT-2 and RoBERTa, but a lot of recent work has developed effective strategies for applying Transformers to images (e.g., ViT \citep{dosovitskiy2021image}) and utilizing pretraining strategies like masked autoencoding (e.g., ViT-MAE \citep{he2022masked}).  \blue{Recent work has explored this topic \citep{zhang2023symbolic}, and it remains an open question} which representation is more effective, and what advantages and disadvantages each representation brings.

\textit{Harmonic vs Temporal}.  What are effective ways to capture both harmonic and temporal information?  The approach described in Section \ref{subsec:tasks_MBCSR} encodes bootleg score columns (or parts thereof) as discrete tokens, and then models temporal information with a Transformer.  But one could alternately encode rectangular blocks of the bootleg score image as discrete tokens, similar to the patches in a ViT model.  In this case, each token would capture both harmonic and temporal information, rather than only harmonic information.

\textit{Raw vs Processed}.  At what level of semantic representation is it best to represent discrete tokens?  On one end of the spectrum, we could represent a bootleg score fragment simply as a sequence of zeros and ones, and then let a Byte Pair Encoder combined with a Transformer model learn the most suitable representation.  On the other end of the spectrum, we could design musically-informed encodings, taking into account domain knowledge such as the split between the left and right hand staves, the fact that staff line positions are cyclical and octave-based, etc.  For example, one could encode a bootleg score column as the text ``C3-G3-E4-C5", which explicitly decomposes the staff line position into octave and class information.  \blue{Recent work has explored this topic \citep{fradet2023byte}, and this remains an open question.}

\textit{Absolute vs Relative}.  How much should the encoding of discrete tokens capture absolute vs relative position information?  One intuitive shortcoming of the Transformer models in Section \ref{subsec:tasks_MBCSR} is that they encode the absolute staff line positions of noteheads, rather than relative position or movement.  In the example given above, the bootleg score column encoded as “C3-G3-E4-C5” could instead be encoded as ``C3-4-5-5" to capture the relative staff line intervals between notes in the chord.  Furthermore, the root of the chord (C3) could itself be expressed relative to a notehead in a previous bootleg score column.

The topics above are issues that could be studied conveniently with the PBSCR data.  Many other similar questions could be rapidly explored, given the simplicity and format of the data.  Thus, even though the dataset does not have complete symbolic score information, it can facilitate rapid exploration of ideas and progress on research questions of broad interest to the MIR community.

\subsection{Data Augmentation}
\label{subsec:questions_dataAug}

Another open research question is, ``What are effective ways to perform data augmentation of symbolic music data?"  The PBSCR data is ideal for exploring data augmentation techniques for two reasons, which we describe below.

First, the PBSCR data has a very simple format.  The fact that the bootleg score is a simple 2D image makes it much easier to explore data augmentation techniques.  For example, many data augmentation strategies developed for computer vision can be applied out-of-the-box with no additional effort, such as cropping, shifting, and MixUp \citep{zhang2017mixup}.  In contrast, for symbolic music formats like MusicXML, it would be much more cumbersome to rapidly explore the same types of data augmentation.  Both existing and novel techniques would be far easier to implement with bootleg score images than with MusicXML data.

Second, the PBSCR data format has several key properties that make such augmentations musically meaningful.  One such property of the bootleg score is the nature of key changes: because the staff line positions (A through G) are cyclical, key changes correspond to simple vertical shifts in the bootleg score representation \blue{(assuming that the key signature is properly adjusted)}.  Similarly, shifts in time correspond to simple horizontal shifts in the bootleg score representation.  The bootleg score also has the property of additivity – if you add a bootleg score event \blue{(i.e.~binary vector of length 62)} describing a chord in the right hand to a bootleg score event describing a left hand chord, the resulting event is simply the sum of the two constituent bootleg score events.  (Note that discrete token-based representations do not have this property.)  It is also worth pointing out that the bootleg score representation is inherently invariant to tempo, since it only captures the sequence of noteheads rather than describing the absolute time between note events.  Because of these properties, operations like cropping and shift and MixUp are very easy to implement and have clear musical interpretations.

The above properties make the PBSCR dataset ideal for exploring data augmentation strategies.  This includes exploring the effectiveness of existing data augmentation techniques from computer vision, as well as quickly implementing and trying domain-specific data augmentation techniques.

\subsection{Integrating Multimodal Information}
\label{subsec:questions_multimodal}

Yet another interesting research question is, ``How can we train models with multiple modalities of data?"  The PBSCR dataset is ideal for exploring this question in MIR for three reasons, which we describe below.

First, the PBSCR data is linked to rich metadata on IMSLP.  In particular, each work in IMSLP has a lot of rich multimodal information: audio recordings, MIDI files, sheet music scores, arrangements and transcriptions, relevant metadata (e.g.,~composer, publisher information, composition date, composer time period, instrumentation), and links to relevant Wikipedia article pages and descriptions (e.g.,~All Music Guide).  Importantly, in keeping with IMSLP’s philosophy, these resources generally have very research-friendly licenses -- most audio recordings and sheet music scores have a Creative Commons license or are in the public domain.

Second, the PBSCR dataset is large enough to study multimodal problems at a nontrivial scale.  Given the size and scale of modern models, it is necessary to have a large enough quantity of data to train large models.  The PBSCR dataset – and certainly IMSLP – fulfills this requirement.  One additional benefit of utilizing IMSLP data is that the website is actively maintained, so the quantity of data will presumably only increase into the future.

Third, the bootleg score representation is well suited for cross-modal and multimodal tasks.  Previous works have demonstrated the effectiveness of the bootleg score representation for cross-modal tasks.  For example, it has been used in cross-modal retrieval to find matches between the Lakh MIDI Dataset and sheet music in IMSLP \citep{yang2021piano}, and it has been used in cross-modal transfer learning to perform composer classification of audio recordings using sheet music as training data \citep{yang2021composer}.  As such, it is well suited to connect multiple representations of music, including sheet music images, symbolic files, and audio.

For these reasons, we believe the PBSCR dataset is particularly well situated to facilitate multimodal research in MIR.  Setting up the infrastructure for specific multimodal tasks is an area for future work.

\section{Conclusion}
\label{sec:concl}

This article motivates, describes, and presents the PBSCR dataset for studying composer recognition of piano sheet music.  Our overarching goal was to create a dataset for studying composer recognition that is ``as accessible as MNIST and as challenging as ImageNet."  To achieve this goal, we use a fixed-length bootleg score representation extracted from piano sheet music images on IMSLP.  This choice allows us to access a large, open, diverse set of data while presenting the data in an extremely simple format that mimics MNIST images.  The dataset itself contains labeled fixed-length bootleg score images for 9-class and 100-class recognition tasks, as well as a large set of variable-length bootleg scores for pretraining.  We include relevant information to connect each bootleg score fragment with the specific work, PDF score, and page from which it was taken, and we scrape, collect, and organize metadata from IMSLP on all works to facilitate multimodal research in the future.  We describe several research tasks that could be studied with the dataset and present baseline results for future works to compare against.  We also discuss open research questions that the PBSCR data is especially well suited to facilitate research on.

\section{Acknowledgments}

This material is based upon work supported by the National Science Foundation under Grant No. 2144050.

	\bibliographystyle{unsrtnat}
	\bibliography{PBSCR.bib}

\begin{thebibliography}{56}
\providecommand{\natexlab}[1]{#1}
\providecommand{\url}[1]{\texttt{#1}}
\expandafter\ifx\csname urlstyle\endcsname\relax
  \providecommand{\doi}[1]{doi: #1}\else
  \providecommand{\doi}{doi: \begingroup \urlstyle{rm}\Url}\fi

\bibitem[Raffel(2016)]{raffel2016learning}
Colin Raffel.
\newblock \emph{Learning-based methods for comparing sequences, with
  applications to audio-to-midi alignment and matching}.
\newblock Columbia University, 2016.

\bibitem[Kong et~al.(2022)Kong, Li, Chen, and Wang]{kong2022giantmidi}
Qiuqiang Kong, Bochen Li, Jitong Chen, and Yuxuan Wang.
\newblock Giantmidi-piano: A large-scale midi dataset for classical piano
  music.
\newblock \emph{Transactions of the International Society for Music Information
  Retrieval}, 5\penalty0 (1):\penalty0 87--98, 2022.

\bibitem[Yang et~al.(2019)Yang, Tanprasert, Jenrungrot, Shan, and
  Tsai]{yang2019midi}
Daniel Yang, Thitaree Tanprasert, Teerapat Jenrungrot, Mengyi Shan, and
  TJ~Tsai.
\newblock {MIDI} passage retrieval using cell phone pictures of sheet music.
\newblock In \emph{Proceedings of the International Society for Music
  Information Retrieval Conference (ISMIR)}, pages 916--923, 2019.

\bibitem[Yang and Tsai(2021{\natexlab{a}})]{yang2021composer}
Daniel Yang and TJ~Tsai.
\newblock Composer classification with cross-modal transfer learning and
  musically-informed augmentation.
\newblock In \emph{Proceedings of the International Society for Music
  Information Retrieval Conference (ISMIR)}, pages 802--809,
  2021{\natexlab{a}}.

\bibitem[Yang and Tsai(2021{\natexlab{b}})]{yang2021piano}
Daniel Yang and TJ~Tsai.
\newblock Piano sheet music identification using dynamic n-gram fingerprinting.
\newblock \emph{Transactions of the International Society for Music Information
  Retrieval}, 4\penalty0 (1), 2021{\natexlab{b}}.

\bibitem[Yang et~al.(2022)Yang, Goutam, Ji, and Tsai]{yang2022large}
Daniel Yang, Arya Goutam, Kevin Ji, and TJ~Tsai.
\newblock Large-scale multimodal piano music identification using marketplace
  fingerprinting.
\newblock \emph{Algorithms}, 15\penalty0 (5):\penalty0 146, 2022.

\bibitem[Tsai and Ji(2020)]{tsai2020composer}
TJ~Tsai and Kevin Ji.
\newblock Composer style classification of piano sheet music images using
  language model pretraining.
\newblock In \emph{Proceedings of the International Society for Music
  Information Retrieval Conference (ISMIR)}, pages 176--183, 2020.

\bibitem[Pape et~al.(2008)Pape, de~Gruijl, and Wiering]{pape2008democratic}
Leo Pape, Jornt de~Gruijl, and Marco Wiering.
\newblock Democratic liquid state machines for music recognition.
\newblock In \emph{Speech, Audio, Image and Biomedical Signal Processing using
  Neural Networks}, pages 191--215. Springer, 2008.

\bibitem[Goienetxea et~al.(2018)Goienetxea, Mendialdua, and
  Sierra]{goienetxea2018use}
Izaro Goienetxea, I{\~n}igo Mendialdua, and Basilio Sierra.
\newblock On the use of matrix based representation to deal with automatic
  composer recognition.
\newblock In \emph{Australasian Joint Conference on Artificial Intelligence},
  pages 531--536. Springer, 2018.

\bibitem[Anan et~al.(2012)Anan, Hatano, Bannai, Takeda, and
  Satoh]{anan2012polyphonic}
Yoko Anan, Kohei Hatano, Hideo Bannai, Masayuki Takeda, and Ken Satoh.
\newblock Polyphonic music classification on symbolic data using dissimilarity
  functions.
\newblock In \emph{Proc. of the International Society for Music Information
  Retrieval Conference (ISMIR)}, pages 229--234, 2012.

\bibitem[Hajj et~al.(2018)Hajj, Filo, and Awad]{hajj2018automated}
Nadine Hajj, Maurice Filo, and Mariette Awad.
\newblock Automated composer recognition for multi-voice piano compositions
  using rhythmic features, n-grams and modified cortical algorithms.
\newblock \emph{Complex \& Intelligent Systems}, 4:\penalty0 55--65, 2018.

\bibitem[Brinkman et~al.(2016)Brinkman, Shanahan, and
  Sapp]{brinkman2016musical}
Andrew Brinkman, Daniel Shanahan, and Craig Sapp.
\newblock Musical stylometry, machine learning and attribution studies: A
  semi-supervised approach to the works of josquin.
\newblock In \emph{Proceedings of the Biennial International Conference on
  Music Perception and Cognition}, pages 91--97, 2016.

\bibitem[Kempfert and Wong(2020)]{kempfert2020does}
Katherine~C Kempfert and Samuel~WK Wong.
\newblock Where does {Haydn} end and {Mozart} begin? {Composer} classification
  of string quartets.
\newblock \emph{Journal of New Music Research}, 49\penalty0 (5):\penalty0
  457--476, 2020.

\bibitem[Herremans et~al.(2016)Herremans, Martens, and
  S{\"o}rensen]{herremans2016composer}
Dorien Herremans, David Martens, and Kenneth S{\"o}rensen.
\newblock Composer classification models for music-theory building.
\newblock \emph{Computational Music Analysis}, pages 369--392, 2016.

\bibitem[McKay and Fujinaga(2006)]{mckay2006jsymbolic}
Cory McKay and Ichiro Fujinaga.
\newblock {jSymbolic}: A feature extractor for {MIDI} files.
\newblock In \emph{Proceedings of the International Computer Music Conference},
  2006.

\bibitem[Hontanilla et~al.(2013)Hontanilla, P{\'e}rez-Sancho, and
  Inesta]{hontanilla2013modeling}
Mar{\'\i}a Hontanilla, Carlos P{\'e}rez-Sancho, and Jose~M Inesta.
\newblock Modeling musical style with language models for composer recognition.
\newblock In \emph{Proceedings of the 6th Iberian Conference on Pattern
  Recognition and Image Analysis}, pages 740--748. Springer, 2013.

\bibitem[Hedges et~al.(2014)Hedges, Roy, and Pachet]{hedges2014predicting}
Thomas Hedges, Pierre Roy, and Fran{\c{c}}ois Pachet.
\newblock Predicting the composer and style of jazz chord progressions.
\newblock \emph{Journal of New Music Research}, 43\penalty0 (3):\penalty0
  276--290, 2014.

\bibitem[Micchi(2018)]{micchi2018neural}
Gianluca Micchi.
\newblock A neural network for composer classification.
\newblock In \emph{Proceedings of the International Society for Music
  Information Retrieval Conference (ISMIR)}, 2018.

\bibitem[Kher(2022)]{kher2022thesis}
Rohit Kher.
\newblock Music composer recognition from midi representation using deep
  learning and n-gram based methods.
\newblock Master's thesis, Dalhousie University, 2022.

\bibitem[Verma and Thickstun(2019)]{verma2019midi}
Harsh Verma and John Thickstun.
\newblock Convolutional composer classification.
\newblock In \emph{Proceedings of the International Society for Music
  Information Retrieval Conference (ISMIR)}, pages 549--556, 2019.

\bibitem[Velarde et~al.(2018)Velarde, Chac{\'o}n, Meredith, Weyde, and
  Grachten]{velarde2018convolution}
Gissel Velarde, Carlos~Cancino Chac{\'o}n, David Meredith, Tillman Weyde, and
  Maarten Grachten.
\newblock Convolution-based classification of audio and symbolic
  representations of music.
\newblock \emph{Journal of New Music Research}, 47\penalty0 (3):\penalty0
  191--205, 2018.

\bibitem[Walwadkar et~al.(2022)Walwadkar, Shatri, Timms, and
  Fazekas]{walwadkar2022compldnet}
Dnyanesh Walwadkar, Elona Shatri, Benjamin Timms, and Gy{\"o}rgy Fazekas.
\newblock {CompldNet}: Sheet music composer identification using deep neural
  network.
\newblock In \emph{Proceedings of the 4th International Workshop on Reading
  Music Systems}, pages 9--14, 2022.

\bibitem[Deepaisarn et~al.(2022)Deepaisarn, Buaruk, Chokphantavee,
  Chokphantavee, Prathipasen, and Sornlertlamvanich]{deepaisarn2022visual}
Somrudee Deepaisarn, Suphachok Buaruk, Sirawit Chokphantavee, Sorawit
  Chokphantavee, Phuriphan Prathipasen, and Virach Sornlertlamvanich.
\newblock Visual-based musical data representation for composer classification.
\newblock In \emph{IEEE International Joint Symposium on Artificial
  Intelligence and Natural Language Processing (iSAI-NLP)}, pages 1--5, 2022.

\bibitem[Kim et~al.(2020)Kim, Lee, Park, Lee, and Choi]{kim2020deep}
Sunghyeon Kim, Hyeyoon Lee, Sunjong Park, Jinho Lee, and Keunwoo Choi.
\newblock Deep composer classification using symbolic representation.
\newblock In \emph{Late-Breaking Demo Session of the International Society for
  Music Information Retrieval Conference}, 2020.

\bibitem[Kong et~al.(2020)Kong, Choi, and Wang]{kong2020large}
Qiuqiang Kong, Keunwoo Choi, and Yuxuan Wang.
\newblock Large-scale midi-based composer classification.
\newblock \emph{arXiv preprint arXiv:2010.14805}, 2020.

\bibitem[Takamoto et~al.(2018)Takamoto, Yoshida, Umemura, and
  Ichikawa]{takamoto2018feature}
Ayaka Takamoto, Mitsuo Yoshida, Kyoji Umemura, and Yuko Ichikawa.
\newblock Feature selection for composer classification method using quantity
  of information.
\newblock In \emph{IEEE International Conference on Knowledge and Smart
  Technology (KST)}, pages 30--33, 2018.

\bibitem[Deepaisarn et~al.(2023)Deepaisarn, Chokphantavee, Chokphantavee,
  Prathipasen, Buaruk, and Sornlertlamvanich]{deepaisarn2023nlp}
Somrudee Deepaisarn, Sirawit Chokphantavee, Sorawit Chokphantavee, Phuriphan
  Prathipasen, Suphachok Buaruk, and Virach Sornlertlamvanich.
\newblock {NLP}-based music processing for composer classification.
\newblock \emph{Scientific Reports}, 13\penalty0 (1):\penalty0 13228, 2023.

\bibitem[Huang and Yang(2020)]{huang2020pop}
Yu-Siang Huang and Yi-Hsuan Yang.
\newblock Pop music transformer: Beat-based modeling and generation of
  expressive pop piano compositions.
\newblock In \emph{Proceedings of the 28th ACM International Conference on
  Multimedia}, pages 1180--1188, 2020.

\bibitem[Hsiao et~al.(2021)Hsiao, Liu, Yeh, and Yang]{hsiao2021compound}
Wen-Yi Hsiao, Jen-Yu Liu, Yin-Cheng Yeh, and Yi-Hsuan Yang.
\newblock Compound word transformer: Learning to compose full-song music over
  dynamic directed hypergraphs.
\newblock In \emph{Proceedings of the AAAI Conference on Artificial
  Intelligence}, pages 178--186, 2021.

\bibitem[Li et~al.(2023)Li, Gong, Chen, and Su]{li2023fine}
Zuchao Li, Ruhan Gong, Yineng Chen, and Kehua Su.
\newblock Fine-grained position helps memorizing more, a novel music compound
  transformer model with feature interaction fusion.
\newblock In \emph{Proceedings of the AAAI Conference on Artificial
  Intelligence}, pages 5203--5212, 2023.

\bibitem[Chou et~al.(2021)Chou, Chen, Chang, Ching, and Yang]{chou2021midibert}
Yi-Hui Chou, I-Chun Chen, Chin-Jui Chang, Joann Ching, and Yi-Hsuan Yang.
\newblock {MidiBERT-piano}: large-scale pre-training for symbolic music
  understanding.
\newblock \emph{arXiv preprint arXiv:2107.05223}, 2021.

\bibitem[Yang et~al.(2021)Yang, Ji, and Tsai]{yang2021deeper}
Daniel Yang, Kevin Ji, and TJ~Tsai.
\newblock A deeper look at sheet music composer classification using
  self-supervised pretraining.
\newblock \emph{Applied Sciences}, 11\penalty0 (4):\penalty0 1387, 2021.

\bibitem[Wo{\l}kowicz and Ke{\v{s}}elj(2013)]{wolkowicz2013evaluation}
Jacek Wo{\l}kowicz and Vlado Ke{\v{s}}elj.
\newblock Evaluation of n-gram-based classification approaches on classical
  music corpora.
\newblock In \emph{International Conference on Mathematics and Computation in
  Music}, pages 213--225, 2013.

\bibitem[Herlands et~al.(2014)Herlands, Der, Greenberg, and
  Levin]{herlands2014machine}
William Herlands, Ricky Der, Yoel Greenberg, and Simon Levin.
\newblock A machine learning approach to musically meaningful homogeneous style
  classification.
\newblock In \emph{Proceedings of the AAAI Conference on Artificial
  Intelligence}, 2014.

\bibitem[Herremans et~al.(2015)Herremans, S{\"o}rensen, and
  Martens]{herremans2015classification}
Dorien Herremans, Kenneth S{\"o}rensen, and David Martens.
\newblock Classification and generation of composer-specific music using global
  feature models and variable neighborhood search.
\newblock \emph{Computer Music Journal}, 39\penalty0 (3):\penalty0 71--91,
  2015.

\bibitem[Saboo et~al.(2015)Saboo, N, and Rajendran]{saboo2015composer}
Krishnakant Saboo, Chaitanya~Prasad N, and Bipin Rajendran.
\newblock Composer classification based on temporal coding in adaptive spiking
  neural networks.
\newblock In \emph{International Joint Conference on Neural Networks (IJCNN)},
  pages 1--8, 2015.

\bibitem[Velarde et~al.(2016)Velarde, Weyde, Chac{\'o}n, Meredith, and
  Grachten]{velarde2016composer}
Gissel Velarde, Tillman Weyde, Carlos Eduardo~Cancino Chac{\'o}n, David
  Meredith, and Maarten Grachten.
\newblock Composer recognition based on 2d-filtered piano-rolls.
\newblock In \emph{Proceedings of the International Society for Music
  Information Retrieval Conference (ISMIR)}, pages 115--121, 2016.

\bibitem[Shuvaev et~al.(2017)Shuvaev, Giaffar, and
  Koulakov]{shuvaev2017representations}
Sergey Shuvaev, Hamza Giaffar, and Alexei~A Koulakov.
\newblock Representations of sound in deep learning of audio features from
  music.
\newblock \emph{arXiv preprint arXiv:1712.02898}, 2017.

\bibitem[Sadeghian et~al.(2017)Sadeghian, Wilson, Goeddel, and
  Olmsted]{sadeghian2017classification}
Pasha Sadeghian, Casey Wilson, Stephen Goeddel, and Aspen Olmsted.
\newblock Classification of music by composer using fuzzy min-max neural
  networks.
\newblock In \emph{Proceedings of the 12th International Conference for
  Internet Technology and Secured Transactions (ICITST)}, pages 189--192, 2017.

\bibitem[Costa and Salazar(2019)]{costa2019dodecaphonic}
Lucas Francesco~Piccioni Costa and Andr{\'e}s Eduardo~Coca Salazar.
\newblock Dodecaphonic composer identification based on complex networks.
\newblock In \emph{2019 8th Brazilian Conference on Intelligent Systems
  (BRACIS)}, pages 765--770, 2019.

\bibitem[Revathi et~al.(2020)Revathi, Vashista, Teja, and
  Nagakrishnan]{revathi2020robust}
A~Revathi, D~Vishnu Vashista, Kuppa Sai~Sri Teja, and R~Nagakrishnan.
\newblock A robust music composer identification system based on cepstral
  feature and models.
\newblock In \emph{Advances in Communication Systems and Networks}, pages
  35--44, 2020.

\bibitem[Foscarin et~al.(2022)Foscarin, Hoedt, Praher, Flexer, and
  Widmer]{foscarin2022concept}
Francesco Foscarin, Katharina Hoedt, Verena Praher, Arthur Flexer, and Gerhard
  Widmer.
\newblock Concept-based techniques for ``musicologist-friendly'' explanations
  in a deep music classifier.
\newblock In \emph{Proceedings of the International Society for Music
  Information Retrieval Conference (ISMIR)}, pages 876--883, 2022.

\bibitem[Simonetta et~al.(2023)Simonetta, Llorens, Serrano,
  Garc{\'\i}a-Portugu{\'e}s, and Torrente]{simonetta2023optimizing}
Federico Simonetta, Ana Llorens, Mart{\'\i}n Serrano, Eduardo
  Garc{\'\i}a-Portugu{\'e}s, and {\'A}lvaro Torrente.
\newblock Optimizing feature extraction for symbolic music.
\newblock In \emph{Proceedings of the International Society for Music
  Information Retrieval Conference (ISMIR)}, 2023.

\bibitem[Zhang et~al.(2023)Zhang, Karystinaios, Dixon, Widmer, and
  Cancino-Chac{\'o}n]{zhang2023symbolic}
Huan Zhang, Emmanouil Karystinaios, Simon Dixon, Gerhard Widmer, and
  Carlos~Eduardo Cancino-Chac{\'o}n.
\newblock Symbolic music representations for classification tasks: A systematic
  evaluation.
\newblock In \emph{Proceedings of the International Society for Music
  Information Retrieval Conference (ISMIR)}, pages 848--858, 2023.

\bibitem[Yang and Tsai(2020)]{yang2020camera}
Daniel Yang and TJ~Tsai.
\newblock Camera-based piano sheet music identification.
\newblock In \emph{Proc. of the International Society for Music Information
  Retrieval Conference (ISMIR)}, pages 481--488, 2020.

\bibitem[Tsai et~al.(2020)Tsai, Yang, Shan, Tanprasert, and
  Jenrungrot]{tsai2020using}
TJ~Tsai, Daniel Yang, Mengyi Shan, Thitaree Tanprasert, and Teerapat
  Jenrungrot.
\newblock Using cell phone pictures of sheet music to retrieve midi passages.
\newblock \emph{IEEE Transactions on Multimedia}, 22\penalty0 (12):\penalty0
  3115--3127, 2020.

\bibitem[Gage(1994)]{gage1994new}
Philip Gage.
\newblock A new algorithm for data compression.
\newblock \emph{C Users Journal}, 12\penalty0 (2):\penalty0 23--38, 1994.

\bibitem[Radford et~al.(2019)Radford, Wu, Child, Luan, Amodei, and
  Sutskever]{radford2019language}
Alec Radford, Jeffrey Wu, Rewon Child, David Luan, Dario Amodei, and Ilya
  Sutskever.
\newblock Language models are unsupervised multitask learners.
\newblock \emph{OpenAI blog}, 1\penalty0 (8):\penalty0 9, 2019.

\bibitem[Kumar et~al.(2022)Kumar, Raghunathan, Jones, Ma, and
  Liang]{kumar2022fine}
Ananya Kumar, Aditi Raghunathan, Robbie Jones, Tengyu Ma, and Percy Liang.
\newblock Fine-tuning can distort pretrained features and underperform
  out-of-distribution.
\newblock In \emph{International Conference on Learning Representations}, 2022.

\bibitem[Liu et~al.(2019)Liu, Ott, Goyal, Du, Joshi, Chen, Levy, Lewis,
  Zettlemoyer, and Stoyanov]{liu2019roberta}
Yinhan Liu, Myle Ott, Naman Goyal, Jingfei Du, Mandar Joshi, Danqi Chen, Omer
  Levy, Mike Lewis, Luke Zettlemoyer, and Veselin Stoyanov.
\newblock {RoBERTa}: A robustly optimized {BERT} pretraining approach.
\newblock \emph{arXiv preprint arXiv:1907.11692}, 2019.

\bibitem[Wu et~al.(2023)Wu, Yu, Tan, and Sun]{wu2023clamp}
Shangda Wu, Dingyao Yu, Xu~Tan, and Maosong Sun.
\newblock {CLaMP}: Contrastive language-music pre-training for cross-modal
  symbolic music information retrieval.
\newblock In \emph{Proceedings of the International Society for Music
  Information Retrieval Conference (ISMIR)}, pages 157--165, 2023.

\bibitem[Radford et~al.(2021)Radford, Kim, Hallacy, Ramesh, Goh, Agarwal,
  Sastry, Askell, Mishkin, Clark, Kureger, and Sutskever]{radford2021learning}
Alec Radford, Jong~Wook Kim, Chris Hallacy, Aditya Ramesh, Gabriel Goh,
  Sandhini Agarwal, Girish Sastry, Amanda Askell, Pamela Mishkin, Jack Clark,
  Gretchen Kureger, and Ilya Sutskever.
\newblock Learning transferable visual models from natural language
  supervision.
\newblock In \emph{International Conference on Machine Learning}, pages
  8748--8763, 2021.

\bibitem[Dosovitskiy et~al.(2021)Dosovitskiy, Beyer, Kolesnikov, Weissenborn,
  Zhai, Unterthiner, Dehghani, Minderer, Heigold, Gelly, Uszkoreit, and
  Houlsby]{dosovitskiy2021image}
Alexey Dosovitskiy, Lucas Beyer, Alexander Kolesnikov, Dirk Weissenborn,
  Xiaohua Zhai, Thomas Unterthiner, Mostafa Dehghani, Matthias Minderer, Georg
  Heigold, Sylvain Gelly, Jakob Uszkoreit, and Neil Houlsby.
\newblock An image is worth 16x16 words: Transformers for image recognition at
  scale.
\newblock In \emph{International Conference on Learning Representations}, 2021.

\bibitem[He et~al.(2022)He, Chen, Xie, Li, Doll{\'a}r, and
  Girshick]{he2022masked}
Kaiming He, Xinlei Chen, Saining Xie, Yanghao Li, Piotr Doll{\'a}r, and Ross
  Girshick.
\newblock Masked autoencoders are scalable vision learners.
\newblock In \emph{Proceedings of the IEEE/CVF Conference on Computer Vision
  and Pattern Recognition (CVPR)}, pages 16000--16009, 2022.

\bibitem[Fradet et~al.(2023)Fradet, Gutowski, Chhel, and Briot]{fradet2023byte}
Nathan Fradet, Nicolas Gutowski, Fabien Chhel, and Jean-Pierre Briot.
\newblock Byte pair encoding for symbolic music.
\newblock In \emph{Proceedings of the Conference on Empirical Methods in
  Natural Language Processing}, pages 2001--2020, 2023.

\bibitem[Zhang et~al.(2018)Zhang, Cisse, Dauphin, and
  Lopez-Paz]{zhang2017mixup}
Hongyi Zhang, Moustapha Cisse, Yann~N Dauphin, and David Lopez-Paz.
\newblock {MixUp}: Beyond empirical risk minimization.
\newblock In \emph{International Conference on Learning Representations}, 2018.

\end{thebibliography}
		
\end{document}